\documentclass[a4paper, 12pt]{article}
\usepackage{jheppub}
\usepackage{amssymb} 
\usepackage{amsmath}
\usepackage{mathtools}
\usepackage{amsfonts}    
\usepackage{dsfont}
\usepackage{young}
\usepackage[vcentermath]{youngtab}
\usepackage[mathscr]{euscript}
\usepackage{tensor}
\usepackage{xcolor}
\usepackage{braket}

\allowdisplaybreaks

\title{Symmetric orbifold OPE from string theory}

\author[]{Vit Sriprachyakul}
\affiliation[]{Institut f\"ur Theoretische Physik,
ETH Z\"urich,\\
Wolfgang-Pauli-Strasse 27,
8093 Z\"urich, Switzerland}
\emailAdd{vsriprachyak@phys.ethz.ch}

\abstract{We discuss how to obtain the symmetric orbifold fusion rule/OPE from the dual string theory. We consider two explicit examples: $k_b=3$ bosonic strings in ${\rm AdS_3}\times X$ in the near-boundary limit and $k=1$ hybrid strings in $\rm AdS_3\times S^3\times\mathbb{T}^4$. The main advantage of these two examples is that they have explicit expressions for the vertex operators in the $x$-basis. We show that the OPE of such vertex operators explicitly captures the longest cycle contribution in the symmetric orbifold fusion rule/OPE. We then argue how one can obtain the shorter-cycle contributions using the screening operators existing in the theories. We also discuss how our general result reduces to the earlier results in the literature.}

\begin{document}

\maketitle

\section{Introduction}

Almost a decade ago, the dual string theory to a free symmetric orbifold theory was identified. The most well-studied example is the duality between the tensionless string theory on $\rm AdS_3\times S^3\times\mathbb{T}^4$ and the symmetric orbifold of $\mathbb{T}^4$, denoted by ${\rm Sym}^N(\mathbb{T}^4)$, \cite{Gaberdiel:2018rqv,Eberhardt:2018ouy,Eberhardt:2019ywk,Dei:2020zui}, see also \cite{Giribet:2018ada,Gaberdiel:2024dva} for the development for the background $\rm AdS_3\times S^3\times S^3\times S^1$. For $\rm AdS_3\times S^3\times\mathbb{T}^4$, many pieces of supporting evidence have been given over the years: the on-shell string spectrum was shown to capture the symmetric orbifold spectrum \cite{Gaberdiel:2018rqv,Eberhardt:2018ouy}, the tree level string amplitudes were argued to reproduce leading order in $1/N$ correlators \cite{Eberhardt:2019ywk,Dei:2020zui,Dei:2023ivl}, extended objects were studied and identified \cite{Gaberdiel:2021kkp,Knighton:2024noc,Gutperle:2024vyp,Harris:2025wak}. The string genus expansion is understood as corresponding to the grand canonical generating function in the dual CFT \cite{Eberhardt:2021jvj,Aharony:2024fid}, see also \cite{Kim:2015gak} for an earlier discussion on this point.
One thing that is still missing on the list is the relation between worldsheet OPE and dual CFT OPE. The formal aspect was discussed in Section 6 of \cite{Eberhardt:2018ouy} and the OPEs between untwisted sector (dual to 1 unit of spectral flow on the worldsheet) operators were studied in \cite{Naderi:2024wqx}. Thus, this leaves the study of the OPEs between nontrivial spectrally flowed vertex operators (dual to twisted sector operators) to be done and this is the main focus of this article.

Given the success of the tensionless limit, it is intriguing to ask if some lessons can be carried over to higher tension. By studying the full $SL(2,\mathbb{R})$ $n$-point functions ($2\leq n\leq 3$), it was argued that the dual CFT to bosonic string theory on ${\rm AdS_3}\times X$ is a deformed symmetric orbifold \cite{Eberhardt:2021vsx}. Subsequently, the study of (full $SL(2,\mathbb{R})$) genus zero 4-point functions in \cite{Dei:2022pkr} further supported the proposed duality. Later in \cite{Knighton:2024qxd,Knighton:2023mhq}, see also \cite{Hikida:2023jyc} for an alternative viewpoint, it was shown using the path integral formalism and adopting the near-boundary limit that the ${\rm AdS_3}\times X$ $n$-point worldsheet amplitudes (for any $n\geq3$) reproduce precisely the perturbative expansion in the dual CFT. This bosonic near-boundary consideration can be supersymmetrised and the dual CFT to superstrings on ${\rm AdS_3}\times X$ was argued to also be a deformed symmetric orbifold \cite{Sriprachyakul:2024gyl,Eberhardt:2021vsx}, see also \cite{Sriprachyakul:2024xih,Yu:2024kxr} for related discussions. The near-boundary limit can also be thought of as describing the long-string sector on the worldsheet \cite{Seiberg:1999xz,Knighton:2024pqh} when $k_b=3$, where $k_b$ is the level of the $SL(2,\mathbb{R})$ WZW model describing $\rm AdS_3$. One of the common features between the tensionless limit and the bosonic string theory in the near-boundary limit is the fact that the spectrally flowed vertex operators, in the so-called \emph{x-basis}, can be written in terms of delta functions. These delta function operators have been proven useful in the correlation function analysis as shown in \cite{Dei:2023ivl,Knighton:2024qxd,Sriprachyakul:2024gyl}. The analysis of the partition function in the near-boundary limit has also been done in \cite{Knighton:2024pqh} for bosonic strings. Thus, for bosonic strings on ${\rm AdS_3}\times X$, the relation between the worldsheet OPE and the symmetric orbifold OPE is also not studied extensively. This will also be addressed in this article. We note that we only expect this relation to manifest itself in a sensible manner for $k_b=3$. This is because, for generic $k_b$, the dual CFT is generally not a free symmetric orbifold theory \cite{Eberhardt:2021vsx,Balthazar:2021xeh,Knighton:2023mhq,Knighton:2024qxd,Sriprachyakul:2024gyl,Chakraborty:2025nlb,Hikida:2023jyc}: we discuss this point in more detail in Section \ref{sec:review ads3}.

In Section \ref{sec:dual CFT OPE from ws}, we consider worldsheet OPEs between nontrivial spectrally flowed states. We show that the worldsheet OPE naturally reproduces the term with the longest cycle length in the symmetric orbifold OPE. We then argue how to account for terms with shorter cycle lengths. The result in this section demonstrates how short-distance behaviours in the spacetime CFT are encoded in the worldsheet theory. We also comment on how our result reduces to the results in \cite{Kutasov:1999xu,Naderi:2024wqx} if we take both of the twists to be one.

The paper is organised as follows. In Section \ref{sec:review ads3} we review necessary material in the bosonic $k_b=3$ ${\rm AdS_3}\times X$ string theory in the near-boundary limit and the tensionless hybrid string theory on $\rm AdS_3\times S^3\times\mathbb{T}^4$. Readers familiar with the aforementioned formulations may wish to skip to Section \ref{sec:dual CFT OPE from ws} which contains our main result. Section \ref{sec:conclusion} contains our summary and some outlook. Appendix \ref{appendix:hybrid convention} summarises our convention for the hybrid description, Appendix \ref{appendix:numerical factor} demonstrates how numerical coefficients in the worldsheet OPEs may be calculated, and Appendix \ref{appendix:proving the claim} provides a justification for the claim used in Section \ref{subsec:shorter length}.

\section{\boldmath Review of string theory on $\rm AdS_3$}\label{sec:review ads3}
In this section, we review conceptually necessary material for the computations in the next section. We will first discuss bosonic string theory on ${\rm AdS_3}\times X$ where $X$ denotes compact directions. Then, we will review the worldsheet theory, using the hybrid description \cite{Berkovits:1999im}, of the tensionless strings on $\rm AdS_3\times S^3\times\mathbb{T}^4$.

\subsection{Bosonic string}
Bosonic string theory on $\rm AdS_3$ can be conveniently described by the $SL(2,\mathbb{R})$ WZW model with level $k_b$ \cite{Maldacena:2000hw,Maldacena:2000kv,Maldacena:2001km}. The spectrum of the theory is constructed from the unitary representations of $\mathfrak{sl}(2,\mathbb{R})$ which are the discrete representations ${\cal D}^+_j$ with $\tfrac{1}{2}<j<\tfrac{k_b-1}{2}$, the continuous representations ${\cal C}_j^\lambda$ with $j\in\tfrac{1}{2}+i\mathbb{R}$ and $\lambda\in[0,1)$, and their spectral flow images \cite{Henningson:1991jc,Hwang:1991ana,Evans:1998qu,Maldacena:2000hw}. The first order action governing the theory is given by \cite{Giveon:1998ns,deBoer:1998gyt,Kutasov:1999xu}
\begin{equation}
S_{\rm AdS_3}=\frac{1}{2\pi}\int d^2z\left( \frac{1}{2}\partial\Phi\bar\partial\Phi-\frac{QR\Phi}{4}+\beta\bar\partial\gamma+\bar\beta\partial\bar\gamma-\nu\beta\bar\beta e^{-Q\Phi} \right)\,.
\label{eq:ads full action}
\end{equation}
Here, $\Phi$ parametrises the radial direction while $\gamma,\bar\gamma$ parametrise the conformal boundary of $\rm AdS_3$, and $\beta,\bar\beta$ are Lagrange multipliers.\footnote{Upon integrating out $\beta,\bar\beta$ and changing variables so that the remaining measure is the Haar measure of $SL(2,\mathbb{R})$, we get back the second order action which is obtained directly from the Polyakov action.} The level $k_b$ of the $SL(2,\mathbb{R})$ WZW model enters the action above through the relation
\begin{equation}
Q=\sqrt{\frac{2}{k_b-2}}\,.
\end{equation}
Note that the interaction term $\beta\bar\beta e^{-Q\Phi}$ vanishes in the near-boundary limit where $\Phi\to+\infty$. Thus, we may replace the full action \eqref{eq:ads full action} by the free action \cite{Giveon:1998ns} whenever we describe the near-boundary worldsheet configuration\footnote{It was also argued in Section 5 of \cite{Sriprachyakul:2024gyl} that, at $k_b=3$, generic long-string spectrally flowed ground state amplitudes do not receive contributions from the interaction term $\beta\bar\beta e^{-Q\Phi}$, at least perturbatively. Hence, if the spectrum only consists of long strings then any ground state amplitudes computed using the free action \eqref{eq:ads free action} is perturbatively exact.}
\begin{equation}
S_{\rm AdS_3, \Phi\to+\infty}=\frac{1}{2\pi}\int d^2z\left( \frac{1}{2}\partial\Phi\bar\partial\Phi-\frac{QR\Phi}{4}+\beta\bar\partial\gamma+\bar\beta\partial\bar\gamma \right)\,.
\label{eq:ads free action}
\end{equation}
In this limit, the $SL(2,\mathbb{R})$ currents can be written in terms of the free fields as follows \cite{Wakimoto:1986gf}
\begin{equation}
\mathcal{J}^+=\beta\,,\quad \mathcal{J}^3=-\frac{1}{Q}\partial\Phi+\beta\gamma\,,\quad \mathcal{J}^-=-\frac{2}{Q}\partial\Phi\,\gamma+\beta\gamma^2-k_b\partial\gamma\,.
\label{eq:currents from Wakimoto fields}
\end{equation}
We note here that the free fields satisfy the OPEs
\begin{equation}
\Phi(z)\Phi(w)\sim -\ln|z-w|^2,\quad \beta(z)\gamma(w)\sim-\frac{1}{z-w}\,.
\label{eq: free field OPEs}
\end{equation}
For completeness, we give the expression of the (left-moving) stress tensor in terms of the Wakimoto fields $\Phi,\beta,\gamma$
\begin{equation}
T_{\rm AdS_3}=-\frac{1}{2}\partial\Phi\partial\Phi-\frac{Q}{2}\partial^2\Phi-\beta\partial\gamma\,.
\end{equation}

So far, our discussion has been generic, however, we will now focus on bosonic strings in $\rm AdS_3$ with $k_b=3$. We will also focus on the long string states and we will give a few reasons why we are interested in this restriction at the end of this subsection. With the aforementioned assumptions, the spectrally flowed vertex operators are given by \cite{Knighton:2023mhq,Knighton:2024qxd}
\begin{equation}
V^w_{m,j,X}(z;x)=e^{\left(\frac{w-1}{\sqrt{2}}-\sqrt{2}ip\right)\Phi}(\partial^w\gamma)^{-m-\frac{1}{2}-ip}\delta_w(\gamma(z)-x)V_X(z)\,,
\label{eq:bosonic tensionless vertex operator}
\end{equation}
where
\begin{equation}
j=\frac{1}{2}+ip\,,
\end{equation}
$w\in\mathbb{Z}^+$, $m$ is the $J^3_0$ eigenvalue of the state \emph{before} the spectral flow, and $V_X$ is the vertex operator from the compact theory $X$. Here, $z$ denotes the worldsheet coordinate whereas $x$ is identified with the boundary coordinate. The delta function $\delta_w(\gamma-x)$ is defined as follows 
\begin{equation}
\delta_w(\gamma-x):=\left(\prod_{i=1}^{w-1}\delta(\partial^i\gamma)\right)\delta(\gamma-x)\,.
\end{equation}
The worldsheet conformal weight of the operator above is
\begin{equation}
\Delta=j(1-j)-hw+\frac{3w^2}{4}+\Delta^X\,,
\end{equation}
where $\Delta^X$ is the conformal weight of $V_X$ and $h$ is the $J^3_0$ eigenvalue of the operator \eqref{eq:bosonic tensionless vertex operator}
\begin{equation}
h=m+\frac{3w}{2}\,.
\end{equation}
For later convenience, we introduce the notation $H$ which is defined to be
\begin{equation}
H=\frac{p^2+\Delta^X}{w}+\frac{3}{4}\left(w-\frac{1}{w}\right)\,,
\end{equation}
for the operator \eqref{eq:bosonic tensionless vertex operator}. Note that $H$ agrees with $h$ whenever the operator \eqref{eq:bosonic tensionless vertex operator} is on-shell, that is,
\begin{equation}
\Delta=1\Leftrightarrow h=H\,,
\end{equation}
and that $H$ is independent of $m$.

Let us now justify why long strings in $k_b=3$ $\rm AdS_3$ string theory in the near-boundary limit are interesting. In \cite{Knighton:2024qxd}, see also \cite{Hikida:2023jyc} for a related discussion, it was shown (for generic $k_b$), by directly evaluating the relevant path integrals, that the near-boundary worldsheet amplitudes reproduce \emph{precisely} the perturbative expansion of a certain deformed symmetric orbifold theory. For a particular worldsheet amplitude, the number of perturbing field insertions $M$ in the dual CFT correlator is dictated by the so-called \emph{j-constraint} \cite{Eberhardt:2019ywk} which reads
\begin{equation}
\sum_{i=1}^{n}j_i-\frac{k_b}{2}(n+2g-2)+(n+3g-3)=-\frac{M}{Q^2}\,.
\label{eq:jconstraint}
\end{equation}
Hence, we see that for $k_b=3$ and for $j_i=\tfrac{1}{2}+ip_i$, we get
\begin{equation}
i\sum_{i=1}^np_i=-\frac{M}{2}\,.
\end{equation}
However, since $M,p_i\in\mathbb{R}$, the above equality implies that
\begin{equation}
M=0
\end{equation}
as well as
\begin{equation}
\sum_{i=1}^np_i=0\,.
\end{equation}
Thus, as long as we consider long string scattering amplitudes at $k_b=3$, the dual CFT correlators are those of the free symmetric orbifold theory. In other words, the long string subsector in the $k_b=3$ $\rm AdS_3$ bosonic string theory should be dual to a free symmetric orbifold theory. It is in this case that we expect to be able to obtain the symmetric orbifold OPE from string theory. Any other values of $k_b$ or the inclusion of short string states would imply that the dual CFT is not a free symmetric orbifold and, in such cases, one should not expect the string theory to contain the information of the OPE of a free symmetric orbifold theory.

\subsubsection{The screening operator: Bosonic strings}
One of the key ingredients of \cite{Knighton:2024qxd} that enters the worldsheet computations is the inclusion of a certain screening operator, the secret representation insertion $\cal O^-$
\begin{equation}
{\cal O}^-=\int d^2z D(z)\bar D(\bar z)\,,
\end{equation}
where 
\begin{equation}
\begin{aligned}
D&=e^{-2\Phi/Q}e^{(k_b-2)\phi+i(k_b-1)\kappa}\\
&=e^{-2\Phi/\sqrt{2}}e^{\phi+2i\kappa}\,,
\label{eq:secret repn}
\end{aligned}
\end{equation}
and similarly for $\bar D$. 
Here, $\phi,\kappa$ are the bosonising fields of $\beta,\gamma$, that is
\begin{equation}
\beta=e^{\phi+i\kappa}\partial(i\kappa),\quad \gamma=e^{-\phi-i\kappa}\,,
\label{eq:bosonisation of beta gamma}
\end{equation}
where the fields $\phi,\kappa$ satisfy the OPEs
\begin{equation}
\phi(z)\phi(w)=\kappa(z)\kappa(w)\sim-\ln(z-w)\,.
\end{equation}
It can be shown that OPEs between $D$ and the currents ${\cal J}^a$ are either regular or, if not regular, a total derivative. Hence, ${\cal O}^-$ is a singlet w.r.t. the $SL(2,\mathbb{R})$ currents. 
We would like to emphasise that the inclusion of the screening operator ${\cal O}^-$ alone is sufficient to show the agreement between (near-boundary) worldsheet amplitudes and (perturbative) spacetime CFT correlators. It is also sufficient to deduce the existence of the twist-2 deformation on the dual CFT side as explained in \cite{Knighton:2024qxd}.

\subsection{Hybrid formalism}
We will now shift gear a bit and review the hybrid description \cite{Berkovits:1999im} of the tensionless strings in $\rm AdS_3\times S^3\times\mathbb{T}^4$, following closely the discussion in \cite{Dei:2023ivl}. The tensionless limit provides an example of holography where both sides of the duality are simultaneously under control. Thus, this offers one of the best playgrounds to explore how the worldsheet theory encodes the information about the symmetric orbifold fusion rule. In the hybrid formalism, the tensionless worldsheet field content consists of
\begin{itemize}
\item $\mathfrak{psu}(1,1|2)_1$ WZW model describing the $\rm AdS_3\times S^3$ part.

\item Topologically twisted $\mathbb{T}^4$ theory.

\item The $\rho,\sigma$ ghosts (and their right-moving analogue).
\end{itemize}
We summarise our convention for the hybrid formalism in Appendix \ref{appendix:hybrid convention}. Following \cite{Dei:2023ivl}, $\mathfrak{psu}(1,1|2)_1$ has a free field realisation in terms of the fields $\beta,\gamma,p_a,\theta^a$, $a=1,2$, satisfying the OPEs
\begin{equation}
\begin{aligned}
\beta(z)\gamma(w)\sim-\frac{1}{z-w},\quad p_a(z)\theta^b(w)\sim\frac{\delta^b_a}{z-w}\,.
\end{aligned}
\end{equation}
Note that $\beta,\gamma$ are bosonic while $p_a,\theta^a$ are fermionic. The currents and the stress tensor are bilinears in these fields, these are given by \eqref{eq:tensionless free field stress tensor} and \eqref{eq:psu currents}. Similar to \eqref{eq:bosonic tensionless vertex operator}, one can write down the vertex operators for the spectrally flowed ground states and they take the form \cite{Dei:2023ivl}
\begin{equation}
V^w_{-2}(z;x)=\exp\left(\frac{w+1}{2}(if_1-if_2)\right)(\partial^w\gamma)^{-m_w}\delta_w(\gamma-x)e^{2\rho+i\sigma+iH}\,,
\label{eq:tensionless hybrid vertex operator with odd w and picture -2}
\end{equation}
for odd $w$ and
\begin{equation}
V^w_{-2}(z;x)=\exp\left(\pm\frac{if_1+if_2}{2}\right)\exp\left(\frac{w+1}{2}(if_1-if_2)\right)(\partial^w\gamma)^{-m^\pm_w}\delta_w(\gamma-x)e^{2\rho+i\sigma+iH}\,,
\label{eq:tensionless hybrid vertex operator with even w and picture -2}
\end{equation}
for even $w$. Here $m_w,m^\pm_w$ are defined as follows
\begin{equation}
m_w=-\frac{(w-1)^2}{4w},\quad m^\pm_w=-\frac{w-2}{4}\,,
\end{equation}
$H$ is the bosonisation of the $\mathbb{T}^4$ fermions, and $f_i,\phi,\kappa$ are the bosonising fields of $\beta,\gamma,p_a,\theta^a$ as explained in Appendix \ref{appendix:hybrid convention}. These vertex operators are (unintegrated) vertex operators in the $(-2)$-picture. For reasons that will become clear in the next section, we also need the vertex operator in the $0$-picture. This is obtained by acting with the picture raising operator (see, for example, eq.(A.11) of \cite{Naderi:2022bus} or eq.(3.4) of \cite{Fiset:2022erp})
\begin{equation}
P=G^+_0(e^{-\rho-iH})_0
\end{equation}
on \eqref{eq:tensionless hybrid vertex operator with odd w and picture -2} and \eqref{eq:tensionless hybrid vertex operator with even w and picture -2}.\footnote{Strictly speaking, we should act the picture raising operator to the vertex operators before the $x$-translation. Explicitly, we have $V_0(z;x)=e^{xJ^+_0}(P\cdot V_{-2})(z;0)e^{-xJ^+_0}$.} The results are
\begin{equation}
V^w_0(z;x)=\exp\left(\frac{w-1}{2}(if_1-if_2)\right)(\partial^w\gamma)^{-m_w+1}\delta_w(\gamma-x)e^{i\sigma}\,,
\label{eq:tensionless hybrid vertex operator with odd w and picture 0}
\end{equation}
for odd $w$ and
\begin{equation}
V^w_0(z;x)=\exp\left(\pm\frac{if_1+if_2}{2}\right)\exp\left(\frac{w-1}{2}(if_1-if_2)\right)(\partial^w\gamma)^{-m^\pm_w+1}\delta_w(\gamma-x)e^{i\sigma}\,,
\label{eq:tensionless hybrid vertex operator with even w and picture 0}
\end{equation}
for even $w$. As briefly mentioned, these are unintegrated vertex operators, they are precisely the gauge-fixed vertex operator $cV_{\Delta=1}$ where $c=e^{i\sigma}$ is the usual conformal ghost. To obtain the integrated vertex operators, we simply strip off the $c$ ghost and then integrate over the worldsheet coordinates. More accurately, we define
\begin{equation}
\Tilde{V}^w_{n}:=G^-_{-1}V^w_n\,,
\label{eq:removing the c ghost}
\end{equation}
where $G^-$ is given in \eqref{eq:full ws twisted N=4}. It can be easily seen that the effect of acting by $G^-_{-1}$ on the physical vertex operators above is to simply remove $e^{i\sigma}$. The integrated vertex operator is therefore
\begin{equation}
\int d^2z \Tilde{V}^w_n(z;x)\Tilde{\bar{V}}^w_{n'}(\bar z;\bar x)\,.
\end{equation}

Note also that the operators \eqref{eq:tensionless hybrid vertex operator with odd w and picture -2} and \eqref{eq:tensionless hybrid vertex operator with even w and picture -2} are physical, see the discussion in Section 3.3 of \cite{Dei:2023ivl}. In particular, the worldsheet conformal weight w.r.t. the ${\cal N}=4$ stress tensor \eqref{eq:full ws twisted N=4} is 0. Furthermore, the $J^3_0$ eigenvalues are
\begin{equation}
h_w=\frac{w^2-1}{4w}\,,
\label{eq:J^3_0 eval for odd w}
\end{equation}
for odd $w$ and 
\begin{equation}
h^\pm_w=\frac{w}{4}\,,
\label{eq:J^3_0 eval for even w}
\end{equation}
for even $w$. These are precisely the twisted sector ground state conformal weights in the dual CFT.

\subsubsection{The screening operator: Hybrid strings}
As argued in \cite{Dei:2023ivl}, it is crucial to add the screening operator that induces poles for $\gamma$ into the worldsheet theory. This operator takes the form
\begin{equation}
\int d^2z D\bar D\,,
\end{equation}
where
\begin{equation}
D=e^{-if_1+if_2}\delta(\beta)=e^{-if_1+if_2}e^{-\phi}\,.
\label{eq:secret repn for hybrid strings}
\end{equation}
In writing the last equality, we assume the bosonisation \eqref{eq:bosonisation of beta gamma}. This screening operator is also a singlet w.r.t. the $SL(2,\mathbb{R})$ currents.

\section{Symmetric orbifold OPE from the worldsheet}\label{sec:dual CFT OPE from ws}
In this section, we argue how to obtain the symmetric orbifold fusion rule/OPE from the dual string theory.

\subsection{Motivation}
Before we dive into the calculations, let us briefly explain the motivation behind the technical manipulation we are about to do. Because of the agreement between worldsheet amplitudes and dual CFT correlators \cite{Dei:2023ivl,Knighton:2024qxd,Sriprachyakul:2024gyl}, which schematically reads 
\begin{equation}
\Braket{\left(\int d^2z D\bar D\right)^N\prod_{i=1}^n\int d^2z_iV^{w_i}_{m_i,j_i,X}(z_i;x_i)}\simeq\Braket{\prod_{i=1}^n{\cal O}^{w_i,p_i,X}(x_i)}\,,
\label{eq:equivalence between ws amp and CFT corr}
\end{equation}
we expect the identification
\begin{equation}
\int d^2zV^w_{m,j,X}(z;x)={\cal O}^{w,p,X}(x)\,.
\label{eq:expected identification}
\end{equation}
Here, ${\cal O}^{w_i,p_i,X}(x_i)$ are operators in the dual CFT.
Thus, if we take the above equality at face value, we would conclude that
\begin{equation}
\begin{aligned}
{\cal O}^{w_1,p_1,X}(x_1){\cal O}^{w_2,p_2,X}(x_2)\sim& \int d^2z_1V^{w_1}_{m_1,j_1,X}(z_1;x_1)\int d^2z_2V^{w_2}_{m_2,j_2,X}(z_2;x_2)\\
\sim&\sum_{l=0}^{{\rm min}(w_1-1,w_2-1)}{\cal O}^{w_1+w_2-1-2l,X}(x_2)+\cdots\\
\sim& \sum_{l=0}^{{\rm min}(w_1-1,w_2-1)}\int d^2z_2V^{w_1+w_2-1-2l}_{X}(z_2;x_2)\,.
\label{eq:expectation}
\end{aligned}
\end{equation}
The second line shows only the single cycle terms in the symmetric orbifold fusion rule \cite{Jevicki:1998bm,Pakman:2009zz,Burrington:2018upk,DeBeer:2019oxm,Ashok:2023kkd,Lunin:2001pw} and $\cdots$ denotes the multi-cycle contributions.
We will shortly see that this expectation is partially correct. In particular, we will see that the worldsheet OPE in \eqref{eq:expectation}, which we define as\footnote{We will see in the next subsection why this is a sensible definition.}
\begin{equation}
\begin{aligned}
&\left(\int d^2z_1V^{w_1}_{m_1,j_1,X}(z_1;x_1)\right)\left(\int d^2z_2V^{w_2}_{m_2,j_2,X}(z_2;x_2)\right)\\
&\hspace{1cm}:=\int d^2z_2\int_{D^\epsilon_{z_2}} d^2z_1V^{w_1}_{m_1,j_1,X}(z_1;x_1)V^{w_2}_{m_2,j_2,X}(z_2;x_2)\,,
\label{eq:def of ws OPE for integrated operators}
\end{aligned}
\end{equation}
where $D^\epsilon_{z_2}$ is a disk of (small) radius $\epsilon$ centred at $z_2$,
will only give rise to the longest single cycle term (the $l=0$ term in the sum) in the symmetric orbifold OPE. This differs from the naive expectation that the fused channel should contain every single cycle contributions. However, there is an apparent reason why the worldsheet OPE between 2 delta function operators above should not give the full CFT fusion rule. Notice that the vertex operators are not the only thing we insert into the worldsheet amplitudes. In the (bosonic) near-boundary consideration \cite{Knighton:2024qxd} as well as in the tensionless limit \cite{Dei:2023ivl}, where the new free field realisation was used, we need to also insert a certain number of the screening operators into the worldsheet correlators as shown in \eqref{eq:equivalence between ws amp and CFT corr}. This insertion is to ensure that the field $\gamma$ has the right singular property to be identified as the covering map. Indeed, we can also see from the Riemann-Hurwitz formula that there should be some interplay from the screening operator.

Consider the following symmetric orbifold correlator
\begin{equation}
\Braket{\sigma_{w_1}(x_1)\sigma_{w_2}(x_2)\cdots}\,,
\label{eq:initial CFT correlator}
\end{equation}
where $\sigma_{w_i}(x_i)$ denotes the (bare) twist field and $\cdots$ stands for operator insertions at other positions. If we were to calculate the correlator above using the Lunin-Mathur method \cite{Lunin:2000yv}, we would have to determine the \emph{covering map}: a map from the covering space to the base space. The Riemann-Hurwitz formula relates the number of singularities $N$ the covering map has to the twists $w_i$. More precisely, we have the relation
\begin{equation}
N=1-g+\frac{1}{2}\left((w_1-1)+(w_2-1)+\sum_{i\geq 3}(w_i-1)\right)\,,
\end{equation}
where $g$ is the genus of the covering surface. Here, we assume that the genus of the base space is zero. To avoid cluttering, we now focus on the sphere covering surface and set $g=0$. The argument we are about to give goes through for any covering surface of fixed genus. Note that if we fuse $\sigma_{w_1}(x_1),\sigma_{w_2}(x_2)$ by sending $x_1\to x_2$, the fusion rule gives
\begin{equation}
\sigma_{w_1}(x_1)\sigma_{w_2}(x_2)\sim\sum_{l=0}^{{\rm min}(w_1-1,w_2-1)}\sigma_{w_1+w_2-1-2l}(x_2)+\cdots\,,
\end{equation}
as we discussed previously.
Thus, the correlator \eqref{eq:initial CFT correlator} becomes
\begin{equation}
\sum_{l=0}^{{\rm min}(w_1-1,w_2-1)}\Braket{\sigma_{w_1+w_2-1-2l}(x_2)\cdots}\,,
\end{equation}
and the Riemann-Hurwitz formula now gives
\begin{equation}
N'=1+\frac{1}{2}\left( (w_1+w_2-1-2l-1)+\sum_{i\geq 3}(w_i-1) \right)\,.
\end{equation}
Hence, we see that only when $l=0$ that we get the equality $N'=N$ while we generally have
\begin{equation}
N'=N-l\,.
\end{equation}
Now, on the worldsheet side, $N'$ is precisely the number of screening operator $D$ one should insert according to the charge conservation law \cite{Knighton:2023mhq,Knighton:2024qxd,Dei:2023ivl}. This means that in order to probe the shorter single cycles in the fusion rule (corresponding to $l>0$), we should have $l$ less screening operators in the worldsheet amplitudes \eqref{eq:equivalence between ws amp and CFT corr} and one natural way to achieve that is to fuse the screening operator to worldsheet OPE \eqref{eq:def of ws OPE for integrated operators}.

\subsection{Longest length single cycle}\label{subsec:longest length}

\subsubsection*{Bosonic strings}
Let us first compute the OPE in bosonic string theory on ${\rm AdS_3}\times X$. Recall that we only consider $k_b=3$ and we assume the near-boundary limit. We will be detailed in this case since an extremely similar manipulation will be useful when discussing hybrid strings. To avoid unnecessary complications, we will only perform the calculation for the left-movers, a completely analogous calculation can, of course, be done for the right-movers. We will also suppress all the numerical factors, these can be easily worked out and we give such a sample calculation in Appendix \ref{appendix:numerical factor}. Consider the physical vertex operator
\begin{equation}
V^w_{m,j,X}(z;x)=e^{(\frac{w-1}{\sqrt{2}}-\sqrt{2}ip)\Phi}(\partial^w\gamma)^{-m-\frac{1}{2}-ip}\delta_w(\gamma(z)-x)V_X(z)\,,
\end{equation}
where $j=\tfrac{1}{2}+ip$.
The OPE between two such vertex operators is
\begin{equation}
\begin{aligned}
&\hspace{-1cm}V^{w_1}_{m_1,j_1,X}(z_1;x_1)V^{w_2}_{m_2,j_2,X}(z_2;x_2)\sim\\
&\hspace{1cm} \sum_{a}z_{12}^{-\Delta^X_1-\Delta^X_2+\Delta^X_a}z_{12}^{-(\frac{w_1-1}{\sqrt{2}}-\sqrt{2}ip_1)(\frac{w_2-1}{\sqrt{2}}-\sqrt{2}ip_2)} \\
&\hspace{2cm}\times e^{(\frac{w_1-1}{\sqrt{2}}-\sqrt{2}ip_1)\Phi(z_1)+(\frac{w_2-1}{\sqrt{2}}-\sqrt{2}ip_2)\Phi(z_2)}V_X^a(z_2)\\
&\hspace{3cm}\times\delta_{w_1}(\gamma(z_1)-x_1)\delta_{w_2}(\gamma(z_2)-x_2)\\
&\hspace{4cm}\times(\partial^{w_1}_{z_1}\gamma)^{-m_1-\frac{1}{2}-ip_1}(\partial^{w_2}_{z_2}\gamma)^{-m_2-\frac{1}{2}-ip_2}\,,
\label{eq:most general OPE}
\end{aligned}
\end{equation}
where we use the $X$ theory OPE (suppressing the structure constant)
\begin{equation}
V^1_X(z_1)V^2_X(z_2)\sim \sum_{a}z_{12}^{-\Delta^X_1-\Delta^X_2+\Delta^X_a}V_X^a(z_2)\,.
\label{eq:compact theory OPE}
\end{equation}
Let us now focus on the terms in \eqref{eq:most general OPE} which are of the form \eqref{eq:bosonic tensionless vertex operator}. There are two-fold benefits for doing so. First, these are vertex operators that we know immediately how to interpret in terms of the dual CFT; and secondly, these are terms that are the easiest to extract from the OPE. We start by simplifying the last 2 lines of \eqref{eq:most general OPE} as follows. We first consider
\begin{equation}
\begin{aligned}
&\delta(\partial_{z_1}\gamma)(\partial^{w_2}_{z_2}\gamma)^{-m_2-\frac{1}{2}-ip_2}\delta_{w_2}(\gamma(z_2)-x_2)\\
&\hspace{0.5cm}=z_{12}^{-(w_2-1)}\delta\left( \frac{\partial_{z_2}^{w_2}\gamma}{(w_2-1)!}+\frac{z_{12}\partial_{z_2}^{w_2+1}\gamma}{w_2!}+... \right)(\partial^{w_2}_{z_2}\gamma)^{-m_2-\frac{1}{2}-ip_2}\delta_{w_2}(\gamma(z_2)-x_2)\\
&\hspace{0.5cm}\sim z_{12}^{-(w_2-1)}(z_{12}\partial^{w_2+1}_{z_2}\gamma)^{-m_2-\frac{1}{2}-ip_2}\delta_{w_2+1}(\gamma(z_2)-x_2)+\cdots\,.
\end{aligned}
\end{equation}
Here $\cdots$ denotes terms that are subleading in $z_{12}:=z_1-z_2$ which contain derivatives of the delta function.
Thus, by repeating this process, we see that
\begin{equation}
\begin{aligned}
&(\partial^{w_1}_{z_1}\gamma)^{-m_1-\frac{1}{2}-ip_1}\prod_{i=1}^{w_1-1}\delta(\partial^i_{z_1}\gamma)(\partial^{w_2}_{z_2}\gamma)^{-m_2-\frac{1}{2}-ip_2}\delta_{w_2}(\gamma(z_2)-x_2)\\
&\hspace{1cm}\sim z_{12}^{-(w_1-1)(w_2-1)-(w_2-1)(m_1+\frac{1}{2}+ip_1)-(w_1-1)(m_2+\frac{1}{2}+ip_2)}\\
&\hspace{2cm}\times(\partial^{w_1+w_2-1}_{z_2}\gamma)^{-(m_1+m_2+\frac{1}{2})-\frac{1}{2}-i(p_1+p_2)}\delta_{w_1+w_2-1}(\gamma(z_2)-x_2)\,.
\end{aligned}
\end{equation}
Lastly, we notice that
\begin{equation}
\begin{aligned}
\delta(\gamma(z_1)-x_1)\delta_{w_1+w_2-1}(\gamma(z_2)-x_2)\sim&\delta\left(x_{12}-\frac{\partial^{w_1+w_2-1}_{z_2}\gamma}{(w_1+w_2-1)!}z_{12}^{w_1+w_2-1}\right)\\
&\hspace{0.5cm}\times\delta_{w_1+w_2-1}(\gamma(z_2)-x_2)+\cdots\,.
\end{aligned}
\end{equation}
Thus, the last 2 lines of \eqref{eq:most general OPE} become (up to some numerical factor)
\begin{equation}
\begin{aligned}
&(\partial^{w_1}_{z_1}\gamma)^{-m_1-\frac{1}{2}-ip_1}\delta_{w_1}(\gamma(z_1)-x_1)(\partial^{w_2}_{z_2}\gamma)^{-m_2-\frac{1}{2}-ip_2}\delta_{w_2}(\gamma(z_2)-x_2)\\
&\hspace{1cm}\sim z_{12}^{-(w_1-1)(w_2-1)-(w_2-1)(m_1+\frac{1}{2}+ip_1)-(w_1-1)(m_2+\frac{1}{2}+ip_2)}\\
&\hspace{2cm}\times(\partial^{w_1+w_2-1}_{z_2}\gamma)^{-(m_1+m_2+\frac{1}{2})-\frac{1}{2}-i(p_1+p_2)}\delta_{w_1+w_2-1}(\gamma(z_2)-x_2)\\
&\hspace{3cm}\times\delta\left(x_{12}-\frac{\partial^{w_1+w_2-1}_{z_2}\gamma}{(w_1+w_2-1)!}z_{12}^{w_1+w_2-1}\right)+\cdots\,.
\end{aligned}
\end{equation}
The OPE \eqref{eq:most general OPE}, where we keep only terms of the form \eqref{eq:bosonic tensionless vertex operator}, is therefore
\begin{equation}
\begin{aligned}
&V^{w_1}_{m_1,j_1,X}(z_1;x_1)V^{w_2}_{m_2,j_2,X}(z_2;x_2)\\
&\hspace{1cm}\sim \sum_az_{12}^{-\Delta_1-\Delta_2+\Delta_a+(w_1+w_2-1)}e^{(\frac{(w_1+w_2-1)-1}{\sqrt{2}}-\sqrt{2}i(p_1+p_2))\Phi(z_2)}V_X^a(z_2)\\
&\hspace{2cm}\times(\partial^{w_1+w_2-1}_{z_2}\gamma)^{-(m_1+m_2+\frac{1}{2})-\frac{1}{2}-i(p_1+p_2)}\delta_{w_1+w_2-1}(\gamma(z_2)-x_2)\\
&\hspace{3cm}\times\delta\left(x_{12}-\frac{\partial^{w_1+w_2-1}_{z_2}\gamma}{(w_1+w_2-1)!}z_{12}^{w_1+w_2-1}\right)\\
&\hspace{1cm}\sim \sum_az_{12}^{-\Delta_1-\Delta_2+\Delta_a+(w_1+w_2-1)}\times(\partial^{w_1+w_2-1}_{z_2}\gamma)\\
&\hspace{2cm}\times V^{w_1+w_2-1}_{m_1+m_2+\frac{3}{2},j_1+j_2-\frac{1}{2},X_a}(z_2;x_2)\times\delta\left(x_{12}-\frac{\partial^{w_1+w_2-1}_{z_2}\gamma}{(w_1+w_2-1)!}z_{12}^{w_1+w_2-1}\right) \,,
\end{aligned}
\end{equation}
where $\Delta_{1,2}$ are the worldsheet conformal weights of $V(z_1),V(z_2)$ and $\Delta_a$ is the worldsheet conformal weight of the operator $V^{w_1+w_2-1}_{m_1+m_2+\frac{3}{2},j_1+j_2-\frac{1}{2},X_a}(z_2;x_2)$, which is defined as in \eqref{eq:bosonic tensionless vertex operator} with $V_X=V_X^a$.
Note that the exponent of $z_{12}$ has the expected form from the standard dimensional analysis of the OPE expansion. Now, we know that $\Delta_{1,2}=1$ by the physical state condition, integrating over $z_1,z_2$ and recalling the definition \eqref{eq:def of ws OPE for integrated operators} gives
\begin{equation}
\begin{aligned}
&\int dz_2dz_1V^{w_1}_{m_1,j_1,X}(z_1;x_1)V^{w_2}_{m_2,j_2,X}(z_2;x_2)\\
&\hspace{1cm}\sim\sum_a x_{12}^{\frac{\Delta_a-1}{w_1+w_2-1}}\int dz_2V^{w_1+w_2-1}_{m_1+m_2+\frac{3}{2}+\frac{\Delta_a-1}{w_1+w_2-1},j_1+j_2-\frac{1}{2},X_a}(z_2;x_2)\\
&\hspace{1cm}\sim\sum_a x_{12}^{H_a-H_1-H_2}\int dz_2V^{w_1+w_2-1}_{m_1+m_2+\frac{3}{2}+\frac{\Delta_a-1}{w_1+w_2-1},j_1+j_2-\frac{1}{2},X_a}(z_2;x_2)\,,
\label{eq:OPE like expansion in the bosonic string theory}
\end{aligned}
\end{equation}
where the second line comes from integrating the delta function over $z_1$
\begin{equation}
\begin{aligned}
\int dz_1\delta\left(x_{12}-\tfrac{\partial^{w_1+w_2-1}_{z_2}\gamma}{(w_1+w_2-1)!}z_{12}^{w_1+w_2-1}\right)=\frac{(w_1+w_2-2)!}{\partial^{w_1+w_2-1}_{z_2}\gamma \times z_{12}^{w_1+w_2-2}}\,,
\end{aligned}
\end{equation}
and we remind the readers that we have used the notation
\begin{equation}
H=\frac{p^2+\Delta^X}{w}+\frac{3}{4}\left(w-\frac{1}{w}\right)\,.
\end{equation}
Note also that the vertex operators on the RHS of \eqref{eq:OPE like expansion in the bosonic string theory} are all on-shell since
\begin{equation}
h_a=\frac{\Delta_a-1}{w_1+w_2-1}+m_1+m_2+\frac{3}{2}+\frac{3}{2}(w_1+w_2-1)=H_a-H_1-H_2+h_1+h_2=H_a\,,
\end{equation}
where we have used $H_{1,2}=h_{1,2}$ which follows from the mass-shell condition. Thus, the worldsheet OPE becomes
\begin{equation}
\begin{aligned}
&\left(\int dz_1V^{w_1}_{m_1,j_1,X}(z_1;x_1)\right)\left(\int dz_2V^{w_2}_{m_2,j_2,X}(z_2;x_2)\right)\\
&\hspace{1cm}\sim\sum_a x_{12}^{h_a-h_1-h_2}\int dz_2V^{w_1+w_2-1}_{h_a-\frac{3}{2}(w_1+w_2-1),j_1+j_2-\frac{1}{2},X_a}(z_2;x_2)+\cdots\,.
\label{eq:OPE expansion in the bosonic string theory}
\end{aligned}
\end{equation}
Quite remarkably, the above worldsheet OPE has the form of the OPE expansion but now in the $x_i$ variables. Furthermore, we see that the exponents of $x_{12}$ are \emph{precisely} the dual CFT conformal weights! Additionally, since all vertex operators in \eqref{eq:OPE expansion in the bosonic string theory} are on-shell, string theory predicts, using \eqref{eq:expected identification}, that the dual symmetric orbifold OPE should be
\begin{equation}
{\cal O}^{w_1,p_1,X}(x_1){\cal O}^{w_2,p_2,X}(x_2)\sim\sum_ax_{12}^{h_a-h_1-h_2}{\cal O}^{w_1+w_2-1,p_1+p_2,X_a}(x_2)+\cdots\,,
\end{equation}
where $\cdots$ denotes the shorter cycle as well as $\rm AdS_3$ excited state contributions. This is a very encouraging piece of evidence that describes how the worldsheet OPEs know about the symmetric orbifold OPEs.
Note also that the delta function constraint gives the relation between the worldsheet and the spacetime separations, that is,
\begin{equation}
x_{12}-\frac{\partial^{w_1+w_2-1}_{z_2}\gamma}{(w_1+w_2-1)!}z_{12}^{w_1+w_2-1}=0\to x_{12}\sim z_{12}^{w_1+w_2-1}\,.
\end{equation}
Before discussing what happens to the shorter cycle length terms, let us now look at the situation in the hybrid tensionless string theory.

\subsubsection*{Hybrid strings}
As far as the computation goes, the manipulation in this case is very similar to the purely bosonic analysis. Thus, we will skip some details which are repetitive. As noted in \cite{Dei:2023ivl} and as reviewed in the previous section, the vertex operator for odd/even spectrally flowed operators are different, see equations \eqref{eq:tensionless hybrid vertex operator with odd w and picture -2}, \eqref{eq:tensionless hybrid vertex operator with even w and picture -2}, \eqref{eq:tensionless hybrid vertex operator with odd w and picture 0}, and \eqref{eq:tensionless hybrid vertex operator with even w and picture 0}. Furthermore, we see that, on the worldsheet, $[w_1]\times[w_2]=[w_1+w_2-1]$, hence, $[\text{odd}]\times[\text{odd}]=[\text{odd}]=[\text{even}]\times[\text{even}]$ whereas $[\text{even}]\times[\text{odd}]=[\text{odd}]\times[\text{even}]=[\text{even}]$. Note also that we now need to take into account the picture number. We will fuse two vertex operators such that one is in the 0-picture and the other is in the $(-2)$-picture. This is purely for convenience. We will first look at the OPE between one odd and one even spectrally flowed vertex operators. We will only present the calculation where the vertex operator with odd spectral flow is in the 0-picture; the case where the even spectrally flowed operator is in the 0-picture should proceed similarly. The OPE we want to compute is (recall the definition \eqref{eq:removing the c ghost})
\begin{equation}
\begin{aligned}
\Tilde{V}_0^{w_1}(z_1;x_1)\Tilde{V}_{-2}^{w_2}(z_2;x_2)\,.
\end{aligned}
\end{equation}
Using \eqref{eq:tensionless hybrid vertex operator with even w and picture -2} and \eqref{eq:tensionless hybrid vertex operator with odd w and picture 0} and keeping only the terms of the form \eqref{eq:tensionless hybrid vertex operator with even w and picture -2}, i.e. the leading terms in $z_{12}$, the OPE is given by
\begin{equation}
\begin{aligned}
&\hspace{-0.3cm}\Tilde{V}_0^{w_1}(z_1;x_1)\Tilde{V}_{-2}^{w_2}(z_2;x_2)\\
&\sim z_{12}^{-\frac{(w_1-1)(w_2-3)}{2}+(w_2-1)(-m_{w_1}+1)-(w_1-1)m^\pm_{w_2}}\times e^{2\rho+iH}\\
&\times\exp\left({\pm\frac{if_1+if_2}{2}}\right)\exp\left( \frac{(w_1+w_2-1)+1}{2}(if_1-if_2) \right)(\partial^{w_1+w_2-1}\gamma)^{-m_{w_1}+1-m^\pm_{w_2}}\\
&\times\delta_{w_1+w_2-1}(\gamma(z_2)-x_2)\times\delta\left(x_{12}-\frac{\partial^{w_1+w_2-1}\gamma}{(w_1+w_2-1)!}z_{12}^{w_1+w_2-1}\right)+\cdots\,.
\end{aligned}
\end{equation}
Integrating over $z_1,z_2$ gives
\begin{equation}
\begin{aligned}
&\hspace{-0.5cm}\int dz_1\Tilde{V}_0^{w_1}(z_1;x_1)\int dz_2\Tilde{V}_{-2}^{w_2}(z_2;x_2)\\
&\sim\int dz_2\exp\left({\pm\frac{if_1+if_2}{2}}\right)(z_2)\exp\left( \frac{(w_1+w_2-1)+1}{2}(if_1-if_2) \right)(z_2)\\
&\hspace{1cm}\times(\partial^{w_1+w_2-1}\gamma)^{m^\pm_{w_1+w_2-1}}\delta_{w_1+w_2-1}(\gamma(z_2)-x_2)\times e^{2\rho+iH}\\
&\hspace{2cm}\times x_{12}^{h^\pm_{w_1+w_2-1}-h_{w_1}-h^\pm_{w_2}}+\cdots\\
&\sim x_{12}^{h^\pm_{w_1+w_2-1}-h_{w_1}-h^\pm_{w_2}}\int dz_2\Tilde{V}_{-2}^{w_1+w_2-1}(z_2;x_2)+\cdots\,,
\label{eq:tensionless odd x even expansion}
\end{aligned}
\end{equation}
where we have used the definitions \eqref{eq:J^3_0 eval for odd w} and \eqref{eq:J^3_0 eval for even w}. The worldsheet OPE again takes the form strikingly similar to the OPE expansion in the dual CFT. Let us now compute the $[\text{odd}]\times[\text{odd}]$ OPE. The OPE in this case reads
\begin{equation}
\begin{aligned}
&\hspace{-0.5cm}\Tilde{V}_0^{w_1}(z_1;x_1)\Tilde{V}_{-2}^{w_2}(z_2;x_2)\\
&\hspace{0.cm}\sim z_{12}^{-\frac{(w_1-1)(w_2-3)}{2}+(w_2-1)(-m_{w_1}+1)-(w_1-1)m_{w_2}}\times e^{2\rho+iH}\\
&\hspace{0.5cm}\exp\left( \frac{(w_1+w_2-1)+1}{2}(if_1-if_2) \right)(\partial^{w_1+w_2-1}\gamma)^{-m_{w_1}+1-m_{w_2}}\\
&\hspace{1cm}\times\delta_{w_1+w_2-1}(\gamma(z_2)-x_2)\times\delta\left(x_{12}-\frac{\partial^{w_1+w_2-1}\gamma}{(w_1+w_2-1)!}z_{12}^{w_1+w_2-1}\right) +\cdots\,.
\end{aligned}
\end{equation}
Integrating over $z_i$ now gives
\begin{equation}
\begin{aligned}
&\hspace{-0.5cm}\int dz_1\Tilde{V}_0^{w_1}(z_1;x_1)\int dz_2\Tilde{V}_{-2}^{w_2}(z_2;x_2)\\
&\sim \int dz_2\exp\left( \frac{(w_1+w_2-1)+1}{2}(if_1-if_2) \right)(\partial^{w_1+w_2-1}\gamma)^{-m_{w_1+w_2-1}}\\
&\hspace{1cm}\times \delta_{w_1+w_2-1}(\gamma(z_2)-x_2)\times e^{2\rho+iH}\times x_{12}^{h_{w_1+w_2-1}-h_{w_1}-h_{w_2}}\\
&\sim x_{12}^{h_{w_1+w_2-1}-h_{w_1}-h_{w_2}}\int dz_2\Tilde{V}_{-2}^{w_1+w_2-1}(z_2;x_2)+\cdots\,.
\label{eq:tensionless odd x odd expansion}
\end{aligned}
\end{equation}
Lastly, let us compute the $[\text{even}]\times[\text{even}]$ OPE. There are three different cases we need to consider. This is because the even spectrally flowed ground operators are two-fold degenerated. Thus, we can either pair up 2 such operators with opposite $K^3_0$ eigenvalues or we can pair up 2 such operators with the same $K^3_0$ eigenvalue. For the sake of clarification, we temporarily introduce the following notation
\begin{equation}
\begin{aligned}
V^{w,\pm}_{-2}(z;x)=&\exp\left(\pm\frac{if_1+if_2}{2}\right)\exp\left(\frac{w+1}{2}(if_1-if_2)\right)(\partial^w\gamma)^{-m^\pm_w}\delta_w(\gamma-x)e^{2\rho+i\sigma+iH}\,,\\
V^{w,\pm}_0(z;x)=&\exp\left(\pm\frac{if_1+if_2}{2}\right)\exp\left(\frac{w-1}{2}(if_1-if_2)\right)(\partial^w\gamma)^{-m^\pm_w+1}\delta_w(\gamma-x)e^{i\sigma}\,,
\label{eq:tensionless hybrid even spectral flow operator with K^3_0 charge refinement}
\end{aligned}
\end{equation}
and we define $\Tilde{V}^{w,\pm}_n(z;x)$ as in \eqref{eq:removing the c ghost}. The pairing between operators with opposite $K^3_0$ eigenvalues gives
\begin{equation}
\begin{aligned}
&\hspace{-0.5cm}\Tilde{V}^{w_1,+}_0(z_1;x_1)\Tilde{V}^{w_2,-}_{-2}(z_2;x_2)\\
&\sim z_{12}^{-\frac{(w_1-1)(w_2-3)+1}{2}+(w_2-1)(-m^\pm_{w_1}+1)-(w_1-1)m^\pm_{w_2}}\times e^{2\rho+iH}\\
&\hspace{0.5cm}\times\exp\left( \frac{(w_1+w_2-1)+1}{2}(if_1-if_2) \right)(\partial^{w_1+w_2-1}\gamma)^{-m^\pm_{w_1}+1-m^\pm_{w_2}}\\
&\hspace{1cm}\times\delta_{w_1+w_2-1}(\gamma(z_2)-x_2)\times\delta\left(x_{12}-\frac{\partial^{w_1+w_2-1}\gamma}{(w_1+w_2-1)!}z_{12}^{w_1+w_2-1}\right)+\cdots \,.
\end{aligned}
\end{equation}
Integrating over $z_i$ gives the expected result
\begin{equation}
\begin{aligned}
&\hspace{-0.5cm}\int dz_1\Tilde{V}^{w_1,+}_0(z_1;x_1)\int dz_2\Tilde{V}^{w_2,-}_{-2}(z_2;x_2)\\
&\sim\int dz_2\exp\left( \frac{(w_1+w_2-1)+1}{2}(if_1-if_2) \right)(\partial^{w_1+w_2-1}\gamma)^{-m_{w_1+w_2-1}}\\
&\hspace{0.5cm}\times\delta_{w_1+w_2-1}(\gamma(z_2)-x_2)\times e^{2\rho+iH}\times x_{12}^{h_{w_1+w_2-1}-h^\pm_{w_1}-h^\pm_{w_2}}+\cdots\\
&\sim x_{12}^{h_{w_1+w_2-1}-h^\pm_{w_1}-h^\pm_{w_2}}\int dz_2\Tilde{V}_{-2}^{w_1+w_2-1}(z_2;x_2)+\cdots\,.
\label{eq:tensionless even+ x even- expansion}
\end{aligned}
\end{equation}
\sloppy Finally, let us compute the OPE $\Tilde{V}^{w_1,+}_0(z_1;x_1)\Tilde{V}^{w_2,+}_{-2}(z_2;x_2)$. The other OPE $\Tilde{V}^{w_1,-}_0(z_1;x_1)\Tilde{V}^{w_2,-}_{-2}(z_2;x_2)$ will not be computed explicitly since its calculation is essentially the same. We will also see that this last OPE differs from the previous OPEs in that it produces an excited state in the leading order as we will now see. Proceeding as done previously, we have
\begin{equation}
\begin{aligned}
&\hspace{-0.5cm}\Tilde{V}^{w_1,+}_0(z_1;x_1)\Tilde{V}^{w_2,+}_{-2}(z_2;x_2)\\
&\sim z_{12}^{-\frac{(w_1-1)(w_2-3)-1}{2}+(w_2-1)(-m^\pm_{w_1}+1)-(w_1-1)m^\pm_{w_2}}\\
&\hspace{0.5cm}\times\exp\left(if_1+if_2\right)\exp\left( \frac{(w_1+w_2-1)+1}{2}(if_1-if_2) \right)(\partial^{w_1+w_2-1}\gamma)^{-m^\pm_{w_1}+1-m^\pm_{w_2}}\\
&\hspace{1cm}\times\delta_{w_1+w_2-1}(\gamma(z_2)-x_2)\delta\left(x_{12}-\frac{\partial^{w_1+w_2-1}\gamma}{(w_1+w_2-1)!}z_{12}^{w_1+w_2-1}\right)\times e^{2\rho+iH}+\cdots\,.
\end{aligned}
\end{equation}
Integrating over $z_i$ gives
\begin{equation}
\begin{aligned}
&\hspace{-0.5cm}\int dz_1\Tilde{V}^{w_1,+}_0(z_1;x_1)\int dz_2\Tilde{V}^{w_2,+}_{-2}(z_2;x_2)\\
&\sim\int dz_2\exp\left(if_1+if_2\right)\exp\left( \frac{(w_1+w_2-1)+1}{2}(if_1-if_2) \right)\times \delta_{w_1+w_2-1}(\gamma(z_2)-x_2)\\
&\hspace{0.5cm}\times(\partial^{w_1+w_2-1}\gamma)^{-m_{w_1+w_2-1}-\frac{1}{w_1+w_2-1}}\times e^{2\rho+iH}x_{12}^{h_{w_1+w_2-1}+\frac{1}{w_1+w_2-1}-h^\pm_{w_1}-h^\pm_{w_2}}+\cdots\,.
\end{aligned}
\end{equation}
Because of the factor $\exp\left(if_1+if_2\right)$, the vertex operator on the RHS does not take the form \eqref{eq:tensionless hybrid vertex operator with odd w and picture -2}. However, the exponent of $x_{12}$ suggests that this vertex operator may correspond to some excitation on the ground state. Indeed, using eq.(3.54) of \cite{Dei:2023ivl}, we see that
\begin{equation}
\begin{aligned}
{\cal J}^{++}_{-\frac{1}{w_1+w_2-1}}\Tilde{V}^{w_1+w_2-1}_{-2}(z_2;0)=&\oint_{z_2} dz p_2(z)\theta^1(z)\gamma^{-\frac{1}{w_1+w_2-1}}(z)\Tilde{V}^{w_1+w_2-1}_{-2}(z_2;0)\\
=&\exp\left(if_1+if_2\right)\exp\left( \frac{(w_1+w_2-1)+1}{2}(if_1-if_2) \right)\\
&\hspace{0.2cm}\times(\partial^{w_1+w_2-1}\gamma)^{-m_{w_1+w_2-1}-\frac{1}{w_1+w_2-1}}\delta_{w_1+w_2-1}(\gamma(z_2))e^{2\rho+iH}\,.
\end{aligned}
\end{equation}
Hence, we see that
\begin{equation}
\begin{aligned}
&\hspace{-0.5cm}\int dz_1\Tilde{V}^{w_1,+}_0(z_1;x_1)\int dz_2\Tilde{V}^{w_2,+}_{-2}(z_2;x_2)\\
&\sim x_{12}^{h_{w_1+w_2-1}+\frac{1}{w_1+w_2-1}-h^\pm_{w_1}-h^\pm_{w_2}}\int dz_2{\cal J}^{++}_{-\frac{1}{w_1+w_2-1}}\Tilde{V}^{w_1+w_2-1}_{-2}(z_2;x_2)+\cdots\,.
\label{eq:tensionless even+ x even+ expansion}
\end{aligned}
\end{equation}
The analysis for $\int dz_1\Tilde{V}^{w_1,+}_0(z_1;x_1)\int dz_2\Tilde{V}^{w_2,+}_{-2}(z_2;x_2)$ simply gives the same result but with the replacement ${\cal J}^{++}\to {\cal J}^{--}$. Note that the unintegrated operator ${\cal J}^{++}_{-1/(w_1+w_2-1)}V^{w_1+w_2-1}_{-2}$ is on-shell since $V^{w_1+w_2-1}_{-2}$ is and ${\cal J}^{++}_{-1/(w_1+w_2-1)}$ is a DDF operator, constructed so that it maps a physical state to a physical state. Furthermore, $h_{w_1+w_2-1}+\tfrac{1}{w_1+w_2-1}$ is the $J^3_0$ eigenvalue of ${\cal J}^{++}_{-1/(w_1+w_2-1)}V^{w_1+w_2-1}_{-2}$ which is to be identified with the spacetime conformal weight for on-shell states. Thus, the worldsheet OPE above again produces an OPE expansion that can be interpreted as the dual CFT OPE expansion.

\subsection{Shorter length single cycle}\label{subsec:shorter length}
In this subsection, we discuss a mechanism for generating lower spectral flow operators in the fused channel corresponding to shorter cycle lengths in the dual symmetric orbifold. As an explicit demonstration of the mechanism, we consider bosonic strings in the near boundary limit; the same mechanism should apply to hybrid strings in order to obtain lower spectral flow operators in the fused channel.

As alluded to in the beginning of this section, it seems the screening operator ${\cal O}^-$ is involved in extracting terms in the worldsheet OPE with smaller spectral flows. In particular, we expect that by fusing \eqref{eq:OPE expansion in the bosonic string theory} (or analogous expressions for hybrid strings) with ${\cal O}^-$, we should be able to probe the shorter cycle contributions in the symmetric orbifold OPE. To proceed, we make the following claim:
\begin{quote}
\textit{equation \eqref{eq:OPE expansion in the bosonic string theory} is valid, up to the order shown, for any spectral flow parameters satisfying $w_1+w_2-1>0$.}
\end{quote}
We emphasise that the statement above is \emph{stronger} than what was derived in Section \ref{subsec:longest length} where we assumed that $w_1,w_2\geq 1$.
We give a heuristic derivation of the above claim in Appendix \ref{appendix:proving the claim}, utilising the idea of Section 3 of \cite{Knighton:2023mhq}.
Next, we note that since ${\cal O}^-$ is a singlet w.r.t. the $SL(2,\mathbb{R})$ currents, we have
\begin{equation}
e^{x{\cal J}^+_0}{\cal O}^-e^{-x{\cal J}^+_0}={\cal O}^-\,.
\end{equation}
In other words, we may $x$-translate the screening operator $D(z;0)$ to $D(z;x)$ without affecting ${\cal O}^-$. For reasons that will be apparent shortly, we will pick the $x$-parameter for $D$ to be $x_1$.
If we now fuse the secret representation insertion $D(z;x_1)$ to the vertex operator on the RHS of \eqref{eq:OPE expansion in the bosonic string theory} using the claim above, we get
\begin{equation}
\begin{aligned}
&\int dzD(z;x_1)\int dz_2V^{w_1+w_2-1}_{h_a-\frac{3}{2}(w_1+w_2-1),j_1+j_2-\frac{1}{2},X_a}(z_2;x_2)\\
&\hspace{1cm}\sim x_{12}^{h'_a-h_a}\int dz_2V^{w_1+w_2-3}_{h'_a-\frac{3}{2}(w_1+w_2-3),j_1+j_2-\frac{1}{2},X_a}(z_2;x_2)+\cdots\,,
\end{aligned}
\end{equation}
where we have used that $\Delta(D)=1,h(D)=0$, $j(D)=\tfrac{k_b-2}{2}=\tfrac{1}{2}$, and $h'_a$ is the $J^3_0$ eigenvalue of the vertex operator $V^{w_1+w_2-3}_{h'_a-\frac{3}{2}(w_1+w_2-3),j_1+j_2-\frac{1}{2},X_a}(z_2;x_2)$ which is on-shell by the same argument in Section \ref{subsec:longest length}. The equation above implies that
\begin{equation}
\begin{aligned}
&{\cal O}^-\left(\int dz_1V^{w_1}_{m_1,j_1,X}(z_1;x_1)\right)\left(\int dz_2V^{w_2}_{m_2,j_2,X}(z_2;x_2)\right)\\
&\hspace{1cm}\sim\sum_a x_{12}^{h_a-h_1-h_2}x_{12}^{h'_a-h_a}\int dz_2V^{w_1+w_2-3}_{h'_a-\frac{3}{2}(w_1+w_2-3),j_1+j_2-\frac{1}{2},X_a}(z_2;x_2)+\cdots\\
&\hspace{1cm}\sim\sum_a x_{12}^{h'_a-h_1-h_2}\int dz_2V^{w_1+w_2-3}_{h'_a-\frac{3}{2}(w_1+w_2-3),j_1+j_2-\frac{1}{2},X_a}(z_2;x_2)+\cdots\,.
\label{eq:OPE expansion in the bosonic string theory for shorter length}
\end{aligned}
\end{equation}
Hence, we see again that, by fusing with the screening operator ${\cal O}^-$, we obtain an expansion that is extremely similar to the expected expansion in the symmetric orbifold theory. As in the previous (longest cycle) case, the exponents of $x_{12}$ have the right values which can also be obtained from the standard scaling argument in the dual CFT and the vertex operators on the RHS of \eqref{eq:OPE expansion in the bosonic string theory for shorter length} are on-shell which implies that the symmetric orbifold OPE should contain the terms
\begin{equation}
\begin{aligned}
{\cal O}^{w_1,p_1,X}(x_1){\cal O}^{w_2,p_2,X}(x_2)\sim&\sum_ax_{12}^{h_a-h_1-h_2}{\cal O}^{w_1+w_2-1,p_1+p_2,X_a}(x_2)\\
&+\sum_ax_{12}^{h'_a-h_1-h_2}{\cal O}^{w_1+w_2-3,p_1+p_2,X_a}(x_2)+\cdots\,,
\end{aligned}
\end{equation}
up to the first sub-leading cycle length. Note that the separations in the middle line of \eqref{eq:OPE expansion in the bosonic string theory for shorter length} combine nicely because of the choice of the $x$-translation we made for $D$. By repeatedly fusing with the screening operator ${\cal O}^-$, we can generate the shorter length contributions in this fashion. Note that by fusing with $l$ ${\cal O}^-$s, the spectral flow of the fused operator is $w_1+w_2-1-2l$. This quantity is always positive for the nontrivial fused channel as we will argue shortly. Hence, we can determine the fusion with ${\cal O}^-$ using \eqref{eq:OPE expansion in the bosonic string theory} since we are guaranteed to respect the bound $w_1+w_2-1-2l>0$. 

We now argue, following \cite{Maldacena:2001km,Eberhardt:2018ouy,Dei:2020zui}, that the lower bound on the spectral flow in the nontrivial fused channel is 
\begin{equation}
{\rm min}(w_1+w_2-1-2l)=|w_1-w_2|+1\,.
\end{equation}
We note that the structure constant in the spacetime CFT OPE can be computed from the following worldsheet ``3-point'' function
\begin{equation}
\Braket{({\cal O}^-)^NV^{w_1}_{m_1,j_1,X}(z_1;x_1)V^{w_2}_{m_2,j_2,X}(z_2;x_2)V^{w_1+w_2-1-2l}_{m_3,j_3,X}(z_3;x_3)}\,.
\end{equation}
Hence, if the above 3-point function vanishes then the fused operator with spectral flow $w_1+w_2-1-2l$ is trivial. Now, note that the degree of the covering map for the above 3-point function is given by
\begin{equation}
N=1+\frac{(w_1-1)+(w_2-1)+(w_1+w_2-1-2l-1)}{2}=w_1+w_2-1-l\,.
\end{equation}
Consistency requires the degree to be at least the maximal value of the spectral flows in the correlator, this restriction yields \cite{Maldacena:2001km,Eberhardt:2018ouy,Dei:2020zui}
\begin{equation}
w_1+w_2-1-l\geq{\rm max}(w_1,w_2,w_1+w_2-1-2l)\,.
\end{equation}
The condition
\begin{equation}
w_1+w_2-1-l\geq w_1+w_2-1-2l
\end{equation}
implies that $l\geq0$ and the remaining condition gives 
\begin{equation}
w_1+w_2-1-l\geq{\rm max}(w_1,w_2)\,,
\end{equation}
which implies
\begin{equation}
l\leq{\rm min}(w_1-1,w_2-1)\,.
\end{equation}
Therefore, we see that
\begin{equation}
|w_1-w_2|+1\leq w_1+w_2-1-2l\leq w_1+w_2-1
\end{equation}
which is exactly the group theoretic bound in the symmetric orbifold theory. Note that we use
\begin{equation}
(w_1+w_2-1)-2\cdot{\rm min}(w_1-1,w_2-1)=|w_1-w_2|+1\,.
\end{equation}

\subsection{Some comments on the agreement with the earlier results in the literature}\label{subsec:agreement with the literature}
Let us first discuss the result of \cite{Kutasov:1999xu}. It can be shown that if the compact CFT $X$ possesses currents $K^a$ satisfying the Kac-Moody algebra
\begin{equation}
K^a(z)K^b(w)\sim\frac{k'\delta^{ab}/2}{(z-w)^2}+\frac{if^{abc}K^c(w)}{z-w}\,,
\end{equation}
then the corresponding worldsheet physical states to the currents $K^a$ are\footnote{To obtain the desired physical worldsheet states, we simply act by $T^a_{-1}$ (see eq.(2.27) of \cite{Giveon:1998ns}) on the $w=1$ spectrally flowed ground state. This $w=1$ ground state is \eqref{eq:bosonic tensionless vertex operator} with $p=0,h=0$ and $V_X=1$.}
\begin{equation}
{\cal K}^a(z;x)=K^a\delta(\gamma-x)\,.
\end{equation}
This vertex operator agrees with eq.(7.4) of \cite{Kutasov:1999xu} as it should. The worldsheet OPE gives
\begin{equation}
\begin{aligned}
{\cal K}(z_1;x_1){\cal K}(z_2;x_2)\sim&\left( 
\frac{k'\delta^{ab}/2}{z_{12}^2}+\frac{if^{abc}K^c}{z_{12}} \right)\delta(\gamma-x_2)\delta(x_{12}-\partial_2\gamma z_{12})+\cdots\,.
\end{aligned}
\end{equation}
Integrating over the worldsheet coordinates yields
\begin{equation}
\begin{aligned}
\int dz_1{\cal K}(z_1;x_1)\int dz_2{\cal K}(z_2;x_2)\sim&\,\frac{k'{\cal I}\delta^{ab}/2}{x_{12}^2}+\frac{if^{abc}{\cal K}^c}{x_{12}}\,,
\end{aligned}
\end{equation}
which is precisely eq.(4.19) of \cite{Kutasov:1999xu}. This gives a nice consistency check of the technique discussed in this section.\footnote{Another OPE that was shown in \cite{Kutasov:1999xu} to coincide with the spacetime calculation is the stress tensor OPE. We have checked that the central charge can be computed correctly using the technique presented in this article.
To correctly determine the subleading poles in the Virasoro OPE using our approach requires a more careful treatment of the subleading terms in the Taylor expansion of $\delta(x_{12}-\partial\gamma z_{12}-\partial^2\gamma z_{12}^2/2-\cdots)=\delta(x_{12}-\partial\gamma z_{12})+\cdots$ which is beyond the scope of this article.}

Next, it was argued in \cite{Naderi:2024wqx} that the delta function $\delta(\gamma-x)$ should be treated as a formal sum. Using this formal treatment, the author was able to show that the dual CFT OPEs can be recovered. We would like to further demonstrate the power of our approach and we will show that the dual CFT OPEs in the tensionless limit can also be reproduced using the procedure we describe in this section. For concreteness, let us consider the OPEs between the spacetime bosons and fermions. The operators corresponding to those fields are \cite{Naderi:2024wqx}\footnote{These vertex operators can be obtained by acting with the DDF operators (eqs.(4.2\&4.10) of \cite{Naderi:2024wqx}), with $r=-1/2$ for fermions and $n=-1$ for bosons, on the $w=1$ spectrally flowed ground state in the 0 picture \eqref{eq:tensionless hybrid vertex operator with odd w and picture 0}.}
\begin{equation}
\begin{gathered}
{\Psi}^{+,i}(z;x)=e^{if_1}\delta(\gamma-x)e^{\rho+iH^i},\quad {\Psi}^{-,i}(z;x)=-e^{-if_2}\delta(\gamma-x)e^{\rho+iH^i}\,,\\
{\partial \bar{{\cal X}}}^{i}(z;x)=\delta(\gamma-x)\partial\bar X^i,\quad {\partial {{\cal X}}}^{i}(z;x)=e^{2if_1-2if_2}(\partial\gamma)^{-2}\delta(\gamma-x)\partial X^ie^{4\rho+2iH}\,.
\end{gathered}
\end{equation}
Note that the operators above are the integrand of integrated vertex operators, thus, there is no ghost factor $e^{i\sigma}$. The fermion OPE is
\begin{equation}
\begin{aligned}
{\Psi}^{+,1}(z_1;x_1){\Psi}^{-,2}(z_2;x_2)=&-\left(e^{if_1}\delta(\gamma-x_1)e^{\rho+iH^1}\right)(z_1)\left(e^{-if_2}\delta(\gamma-x_2)e^{\rho+iH^2}\right)(z_2)\\
\sim&-\frac{1}{z_{12}}e^{if_1-if_2}e^{2\rho+iH}\delta(\gamma(z_2)-x_2)\delta(x_{12}-\partial_2\gamma z_{12})\,.
\end{aligned}
\end{equation}
Upon integrating over $z_{1,2}$, we get
\begin{equation}
\begin{aligned}
\int dz_1{\Psi}^{+,1}(z_1;x_1)\int dz_2{\Psi}^{-,2}(z_2;x_2)\sim&-\int dz_2\frac{1}{z_{12}}e^{if_1-if_2}e^{2\rho+iH}\delta(\gamma(z_2)-x_2)\times\frac{1}{\partial_2\gamma}\\
\sim&-\frac{1}{x_{12}}\int dz_2 e^{if_1-if_2}e^{2\rho+iH}\delta(\gamma(z_2)-x_2)\,.
\end{aligned}
\end{equation}
The vertex operator in the last line is simply the $w=1$ ground state (in the $(-2)$-picture) which corresponds to the untwisted ground state in the dual CFT. This string theory OPE reproduces exactly the expected spacetime OPE which is
\begin{equation}
\psi^{+,1}(x_1)\psi^{-,2}(x_2)\sim-\frac{1}{x_{12}}\,.
\end{equation}
The computation for the bosonic OPE is very similar, in this case, we obtain
\begin{equation}
\begin{aligned}
&\hspace{-1cm}{\partial \bar{{\cal X}}}^{i}(z_1;x_1){\partial {{\cal X}}}^{j}(z_2;x_2)\\
=&\left(\delta(\gamma-x_1)\partial\bar X^i\right)(z_1)\left(e^{2if_1-2if_2}(\partial\gamma)^{-2}\delta(\gamma-x_2)\partial X^je^{4\rho+2iH}\right)(z_2)\\
\sim&\frac{\delta^{ij}}{z_{12}^2}e^{2if_1-2if_2}(\partial\gamma)^{-2}\delta(\gamma-x_2)\partial X^je^{4\rho+2iH}\delta(x_{12}-\partial_2\gamma z_{12})\,.
\end{aligned}
\end{equation}
Integrating over the $z_i$, we get
\begin{equation}
\begin{aligned}
&\hspace{-1cm}\int dz_1{\partial \bar{{\cal X}}}^{i}(z_1;x_1)\int dz_2{\partial {{\cal X}}}^{j}(z_2;x_2)\\
\sim&\frac{\delta^{ij}}{x_{12}^2}\int dz_2e^{2if_1-2if_2}(\partial\gamma)^{-1}\delta(\gamma-x_2)\partial X^je^{4\rho+2iH}\,.
\end{aligned}
\end{equation}
The operator on the RHS above is just the $w=1$ spectrally flowed ground state in the $(-4)$-picture as noted in \cite{Naderi:2024wqx}. Again, this reproduces the expected spacetime OPE. 

Therefore, we see that our approach described in this section can be used to compute more general OPEs as we just showed. Furthermore, by carefully keeping track of the signs and numerical factors, we are able to precisely reproduce the spacetime OPEs as we have seen. Similarly, one can in principle recover the OPEs in Section 5 of \cite{Naderi:2024wqx} using our approach and we have checked that this is indeed the case. Note that we do not need to worry about the shorter cycle contributions in this subsection because the symmetric orbifold fusion rule between at least one untwisted sector field only contains the longest cycle contribution. This can also be seen from the fact that $w_1+w_2-1=w_2=|w_1-w_2|+1$ where we have used that $w_1=1$.

\section{Conclusion and outlook}\label{sec:conclusion}
We have derived and shown (see \eqref{eq:OPE expansion in the bosonic string theory}, \eqref{eq:tensionless odd x even expansion}, \eqref{eq:tensionless odd x odd expansion}, \eqref{eq:tensionless even+ x even- expansion}, and \eqref{eq:tensionless even+ x even+ expansion}) that the OPE between two delta function vertex operators reproduces the longest cycle contribution in the symmetric orbifold OPE, up to numerical coefficients which can be worked out as explained in Appendix \ref{appendix:numerical factor}. We have mostly restricted ourselves to the vertex operators of the form \eqref{eq:bosonic tensionless vertex operator}, \eqref{eq:tensionless hybrid vertex operator with odd w and picture -2}, and \eqref{eq:tensionless hybrid vertex operator with even w and picture -2} in the fused channel.\footnote{This does not necessarily mean that we keep only the leading order in the worldsheet separation ($z_{12}$) in the bosonic string theory since there are contributions from operators with non-minimal conformal weight from the compact theory $X$, see \eqref{eq:most general OPE}. However, in the tensionless limit, this does mean we consider the leading order in the worldsheet separation.} We also discuss a mechanism for generating shorter-cycle contributions which is to make use of the secret representation insertion (screening operator) as explained in Section \ref{subsec:shorter length}. Furthermore, we demonstrate how our approach leads to results which agree with the earlier results in the literature, see Section \ref{subsec:agreement with the literature}. Our calculations shed some light on how the worldsheet OPEs and the spacetime OPEs are related.

Note that our calculations imply that the naive relation like \eqref{eq:expectation} is not entirely correct in the free field description of the worldsheet theory. Thus, some corrections must be made. If we assume that the mapping \eqref{eq:expected identification}
remains unchanged, then the most natural modification is to change the prescription \eqref{eq:expectation} to (schematically)
\begin{equation}
\begin{aligned}
&e^{{\cal O}^-}\left(\int d^2z_1V^{w_1}_{m_1,j_1,X}(z_1;x_1)\right)\left(\int d^2z_2V^{w_2}_{m_2,j_2,X}(z_2;x_2)\right)\\
&\hspace{1cm}\sim {\cal O}^{w_1,p_1,X}(x_1){\cal O}^{w_2,p_2,X}(x_2)\,.
\label{eq:proposed identification}
\end{aligned}
\end{equation}
This proposal solves two problems in one go.  First it preserves the AdS/CFT mapping \eqref{eq:expected identification} which we believe is a natural identification. Second, $e^{{\cal O}^-}$ supplies the worldsheet OPE with the necessary factors of ${\cal O}^-$ to generate the shorter cycle contributions. Furthermore, $e^{{\cal O}^-}$ also appears in the integrand of the worldsheet path integral, thus, its appearance in \eqref{eq:proposed identification} is somewhat natural. We currently do not have solid evidence for this claim beyond what we already mentioned, but it would be good to see whether this proposal is correct or not. We leave the investigation to future work.

There are a few concrete future directions that are worth exploring given what we have done in this article.

\subsubsection*{Systematising the descendants}
As mentioned in the conclusion, we only focus on specific terms in the fused channel and it would be desirable to understand how the contributions we have neglected organise themselves. We anticipate that they should combine into vertex operators that can be written as DDF operators acting on operators of the form \eqref{eq:bosonic tensionless vertex operator}, \eqref{eq:tensionless hybrid vertex operator with odd w and picture -2}, and \eqref{eq:tensionless hybrid vertex operator with even w and picture -2}. We have already seen an example of this at the leading order, say, in \eqref{eq:tensionless even+ x even+ expansion}. Also, there were some conjectures \cite{Burrington:2018upk,DeBeer:2019oxm} that the bare twist field OPEs only generate (fractional) Virasoro (or its ${\cal N}=4$ version, depending on the amount of supersymmetry in the dual CFT) descendants. It would be interesting to explore this conjecture from the worldsheet point of view. Eventually, it would also be crucial to understand the numerical coefficients we ignored and see if the worldsheet OPE can \emph{exactly} compute the symmetric orbifold OPE. For example, perhaps using \eqref{eq:numerical prefactor BEFORE integration} and \eqref{eq:numerical prefactor AFTER integration}, can the dual string theory reproduce the structure constant predicted, say, in \cite{Lunin:2000yv,Burrington:2018upk,Lunin:2001pw}.

\subsubsection*{\boldmath Understanding OPEs in the full $\mathfrak{sl}(2,\mathbb{R})_3$ theory}
So far, we have dealt with the $\mathfrak{sl}(2,\mathbb{R})_3$ WZW CFT using the Wakimoto (free field) realisation. This formulation requires screening operators for the worldsheet correlation functions to be non-vanishing. It is intriguing to see what changes if we were to treat the $\mathfrak{sl}(2,\mathbb{R})_3$ WZW formally without using the free field (Wakimoto) description. Firstly, we would expect the worldsheet OPE to reproduce \emph{all} possible single cycle contributions, not just the longest one, since there are now no screening operators. Secondly, it would be good to understand the mechanism that relates the worldsheet separation ($z_{12}$) to the spacetime separation ($x_{12}$). In the Wakimoto realisation, this is achieved by imposing the delta function constraint. Clearly, there should be an analogue of that constraint/relation in the formal treatment and it would be interesting to understand that in more detail.

\subsubsection*{Symmetric orbifold fusion rules as line operator fusion rules}
As was argued in \cite{Knighton:2023mhq}, the spectral flow action can be generated by wrapping a certain non-local line operator around an unflowed vertex operator. Indeed, we have made an initial use of this idea in Appendix \ref{appendix:proving the claim} in order to heuristically justify the claim in Section \ref{subsec:shorter length} and it would be compelling to explore the consequences of this idea further. In this approach, the question of computing the worldsheet OPE then becomes the question of understanding (mainly) the fusion of the line operators. What's more, both signs of spectral flow can be treated on equal footing and the form of the spectrally flowed vertex operators is uniform regardless of the sign of the spectral flow. It is intriguing to see whether this could be used to derive the worldsheet fusion rule in the $x$-basis for generic $w_1,w_2\in\mathbb{Z}$. 

\acknowledgments
We thank Matthias Gaberdiel, Anthony Houppe, Bob Knighton, Ji Hoon Lee, Edward Mazenc, Kiarash Naderi, and Beat Nairz for many stimulating discussions and Matthias Gaberdiel and Beat Nairz for invaluable comments on the draft version of this article. We are grateful to Matthias Gaberdiel for questions and discussions that initiated this project and for his encouragement during the final stage of this work. The work of VS is supported by a grant from the Swiss National Science Foundation, and the work of the group at ETH is also supported in part by the Simons Foundation grant 994306  (Simons Collaboration on Confinement and QCD Strings), as well as the NCCR SwissMAP that is also funded by the Swiss National Science Foundation. 

\appendix

\section{Hybrid formalism convention}\label{appendix:hybrid convention}
In this appendix, we fix our convention for the hybrid formalism used in the main text. We largely follow the convention of \cite{Dei:2023ivl} which we review here for completeness. We will also focus on the left movers since the discussion for the right movers is completely analogous. In the hybrid description of the tensionless strings on $\rm AdS_3\times S^3\times\mathbb{T}^4$, the worldsheet theory possesses a small ${\cal N}=4$ algebra
\begin{equation}
\begin{gathered}
T=T_{\text{free}}-\frac{1}{2}((\partial\rho)^2+(\partial\sigma)^2)+\frac{3}{2}\partial^2(\rho+i\sigma)+T_C\,,\\
G^+=e^{-\rho}Q+e^{i\sigma}T-\partial(e^{i\sigma}\partial(\rho+iH))+G^+_C,\quad G^-=e^{-i\sigma}\,,\\
\Tilde{G}^-=e^{-2\rho-i\sigma-iH}Q-e^{-\rho-iH}T-e^{-\rho-i\sigma}\Tilde{G}^-_C+e^{-\rho-iH}\left( i\partial\sigma\partial(\rho+iH)+\partial^2(\rho+iH) \right)\,,\\
\Tilde{G}^+=e^{\rho+iH},\quad J=\frac{1}{2}\partial(\rho+i\sigma+iH),\quad J^{\pm\pm}=e^{\pm(\rho+i\sigma+iH)}\,,
\label{eq:full ws twisted N=4}
\end{gathered}
\end{equation}
where the generators with the subscript $C$ are the generators in the topologically twisted $\mathbb{T}^4$ theory and 
\begin{equation}
Q=p_1p_2\partial\gamma\,.
\end{equation}
The stress tensor $T_{\text{free}}$ is given by
\begin{equation}
T_{\text{free}}=-\beta\partial\gamma-p_a\partial\theta^a\,.
\label{eq:tensionless free field stress tensor}
\end{equation}
Note that the ghosts satisfy the OPE
\begin{equation}
\rho(z)\rho(w)=\sigma(z)\sigma(w)\sim-\ln(z-w)\,.
\end{equation}
For the $\mathbb{T}^4$ theory, the convention is as follows. Before the topological twist, the fields $X^i,\bar{X}^i,\psi^{\alpha,i}$ where $i=1,2$ and $\alpha=\pm$ have weight 0, 0, and $\tfrac{1}{2}$ respectively. These fields satisfy the OPEs
\begin{equation}
X^i(z)\bar{X}^j(w)\sim\delta^{ij}\ln|z-w|^2,\quad \psi^{\alpha,i}(z)\psi^{\beta,j}(w)\sim-\frac{\epsilon^{\alpha\beta}\epsilon^{ij}}{z-w}\,,
\end{equation}
where $\epsilon^{+-}=\epsilon^{12}=1$. The ${\cal N}=4$ generators in the $\mathbb{T}^4$ theory are given by
\begin{equation}
\begin{gathered}
T'_C=\partial X^i\partial\bar{X}^i+\frac{1}{2}\epsilon^{\alpha\beta}\epsilon^{ij}\psi^{\alpha,i}\psi^{\beta,j},\quad J'_C=-\frac{1}{4}|\epsilon^{\alpha\beta}|\epsilon^{ij}\psi^{\alpha,i}\psi^{\beta,j}\,,\\
(J')^{\pm\pm}_C=\pm\psi^{\pm,1}\psi^{\pm,2},\quad (G')^+_C=\partial\bar{X}^i\psi^{+,i},\quad (G')^-_C=-\epsilon^{ij}\partial{X}^i\psi^{-,j}\,,\\
\Tilde{G'}^+_C=-\epsilon^{ij}\partial{X}^i\psi^{+,j},\quad \Tilde{G'}^-_C=-\partial\bar{X}^i\psi^{-,i}\,,
\label{eq:T4 N=4}
\end{gathered}
\end{equation}
where the prime is inserted to emphasise that these are generators before the twist. After the topological twist, we simply add $\partial J'_C$ to $T'_C$, that is, we have
\begin{equation}
\begin{gathered}
T_C=\partial X^i\partial\bar{X}^i+\frac{1}{2}\epsilon^{\alpha\beta}\epsilon^{ij}\psi^{\alpha,i}\psi^{\beta,j}+\partial J_C,\quad J_C=-\frac{1}{4}|\epsilon^{\alpha\beta}|\epsilon^{ij}\psi^{\alpha,i}\psi^{\beta,j}\,,\\
J^{\pm\pm}_C=\pm\psi^{\pm,1}\psi^{\pm,2},\quad G^+_C=\partial\bar{X}^i\psi^{+,i},\quad G^-_C=-\epsilon^{ij}\partial{X}^i\psi^{-,j}\,,\\
\Tilde{G}^+_C=-\epsilon^{ij}\partial{X}^i\psi^{+,j},\quad \Tilde{G}^-_C=-\partial\bar{X}^i\psi^{-,i}\,,
\label{eq:T4 twisted N=4}
\end{gathered}
\end{equation}
as the generators of the topologically twisted ${\cal N}=4$ algebra for the $\mathbb{T}^4$ theory. Furthermore, the bosonisation of the fermions is given by
\begin{equation}
\begin{gathered}
\psi^{+,1}=e^{iH^1},\quad \psi^{+,2}=e^{iH^2},\quad \psi^{-,1}=e^{-iH^2},\quad \psi^{-,2}=-e^{-iH^1}\,,
\label{eq:T4 bosonisation}
\end{gathered}
\end{equation}
where
\begin{equation}
H^i(z)H^j(w)\sim-\delta^{ij}\ln(z-w)\,.
\end{equation}
By straightforward calculations, we see that
\begin{equation}
J_C=\frac{1}{2}\partial (iH)\,,
\end{equation}
where
\begin{equation}
H:=H^1+H^2\,.
\end{equation}
Let us now turn to describe the $\mathfrak{psu}(1,1|2)_1$ part of the worldsheet theory. The $\mathfrak{psu}(1,1|2)_1$ currents are mostly bilinears in the free fields, explicitly the currents are given by
\begin{equation}
\begin{gathered}
J^+=\beta,\quad J^3=\beta\gamma+\frac{1}{2}p_a\theta^a,\quad J^-=((\beta\gamma)\gamma)+(p_a\theta^a)\gamma=\beta\gamma^2-\partial\gamma+(p_a\theta^a)\gamma\,,\\
K^+=p_2\theta^1,\quad K^3=-\frac{1}{2}p_1\theta^1+\frac{1}{2}p_2\theta^2,\quad K^-=p_1\theta^2\,,\\
S^{+++}=p_2,\quad S^{+-+}=p_1,\quad S^{-++}=-\gamma p_2,\quad S^{---}=((\beta\gamma+p_a\theta^a)\theta^2)\,,\\
S^{++-}=\beta\theta^1,\quad S^{+--}=-\beta\theta^2,\quad S^{--+}=-\gamma p_1,\quad S^{-+-}=-((\beta\gamma+p_a\theta^a)\theta^1)\,.
\label{eq:psu currents}
\end{gathered}
\end{equation}
Note that there are two definitions of normal ordering in the expressions above. One is denoted by the round bracket $(...)$ and is defined as
\begin{equation}
(AB)(w):=\oint_{w}dz \frac{A(z)B(w)}{z-w}\,,
\end{equation}
and the other is the conformal normal ordering, as defined in Section 2 of \cite{Polchinski:1998rq},\footnote{One may also consult a brief review in Appendix C of \cite{Gaberdiel:2022als} where the differential contraction operators were given for the $bc$ system and the chiral boson system, see eqs.(C.14\& C.16).} which is implicitly implied whenever we write the product of fields without any brackets. It is useful to bosonise these free fields, we follow \cite{Dei:2023ivl} and bosonise them as
\begin{equation}
\begin{gathered}
\beta=e^{\phi+i\kappa}\partial(i\kappa),\quad \gamma=e^{-\phi-i\kappa},\quad \theta^1=e^{if_1},\quad p_1=e^{-if_1},\quad \theta^2=e^{-if_2},\quad p_2=e^{if_2}\,,
\label{eq:bosonisation of psu fields}
\end{gathered}
\end{equation}
where
\begin{equation}
\phi(z)\phi(w)=\kappa(z)\kappa(w)\sim-\ln(z-w),\quad f_i(z)f_j(w)\sim-\delta_{ij}\ln(z-w)\,.
\end{equation}
In terms of these bosonising fields, the stress tensor $T_{\text{free}}$ becomes
\begin{equation}
\begin{aligned}
T_{\text{free}}=&-\frac{1}{2}((\partial\phi)^2+(\partial\kappa)^2)+\frac{1}{2}\partial^2(\phi+i\kappa)\\
&\hspace{1cm}-\frac{1}{2}\partial f_i\partial f_i+\frac{1}{2}\partial^2(if_1)-\frac{1}{2}\partial^2(if_2)\,.
\end{aligned}
\end{equation}
With respect to the total stress tensor in \eqref{eq:full ws twisted N=4}, the exponential operators have conformal weights
\begin{equation}
\begin{gathered}
\Delta(e^{imf_j})=\frac{m^2+(-)^jm}{2},\quad \Delta(e^{m\phi})=\frac{-m^2+m}{2},\quad \Delta(e^{im\kappa})=\frac{m^2-m}{2}\,,\\
\Delta(e^{m\rho})=\frac{m(3-m)}{2},\quad \Delta(e^{im\sigma})=\frac{m(m-3)}{2}\,,\\
\Delta(e^{iH})=0,\quad \Delta(e^{-iH})=2\,.
\end{gathered}
\end{equation}

\section{Computing numerical prefactors}\label{appendix:numerical factor}
In the main text, we did not explicitly compute the numerical prefactor of the leading term in the OPE
\begin{equation}
\begin{aligned}
(\partial^{w_1}_1\gamma)^{m_1}\delta_{w_1}(\gamma(z_1)-x_1)(\partial^{w_2}_2\gamma)^{m_2}\delta_{w_2}(\gamma(z_2)-x_2)\,,
\end{aligned}
\end{equation}
since we were primarily interested in the short distance behaviour in the $x_{12}$ variable. We will now do so in this appendix. As in the main text, we start by considering the OPE
\begin{equation}
\begin{aligned}
&\delta(\partial_1\gamma)(\partial_2^{w_2}\gamma)^{m_2}\delta_{w_2}(\gamma(z_2)-x_2)\\
&\hspace{1cm}\sim(-)^{m_2}(w_2-1)!w_2^{-m_2}z_{12}^{-(w_2-1)+m_2}(\partial_2^{w_2+1}\gamma)^{m_2}\delta_{w_2+1}(\gamma(z_2)-x_2)+\cdots\,.
\end{aligned}
\end{equation}
Repeating the process, we get
\begin{equation}
\begin{aligned}
&(\partial^{w_1}_1\gamma)^{m_1}\delta_{w_1}(\gamma(z_1)-x_1)(\partial^{w_2}_2\gamma)^{m_2}\delta_{w_2}(\gamma(z_2)-x_2)\\
&\hspace{0.5cm}\sim(-)^{(w_1-1)m_2}[(w_2-1)!]^{w_1-1-m_1}w_2^{-(w_1-1)m_2}z_{12}^{-(w_1-1)(w_2-1)+(w_1-1)m_2+(w_2-1)m_1}\\
&\hspace{1cm}\times(\partial_2^{w_1+w_2-1}\gamma)^{m_1+m_2}\delta_{w_1+w_2-1}(\gamma(z_2)-x_2)\\
&\hspace{1.5cm}\times\delta\left( x_{12}-z_{12}^{w_1+w_2-1}\frac{\partial^{w_1+w_2-1}_2\gamma}{(w_1+w_2-1)!} \right)+\cdots\,.
\label{eq:numerical prefactor BEFORE integration}
\end{aligned}
\end{equation}
Note that integrating over $z_1$ gives
\begin{equation}
\begin{aligned}
&\hspace{-0.5cm}\int dz_1(\partial^{w_1}_1\gamma)^{m_1}\delta_{w_1}(\gamma(z_1)-x_1)(\partial^{w_2}_2\gamma)^{m_2}\delta_{w_2}(\gamma(z_2)-x_2)\\
&\hspace{0.5cm}\sim(-)^{(w_1-1)m_2}[(w_2-1)!]^{w_1-1-m_1}w_2^{-(w_1-1)m_2}(w_1+w_2-2)!\\
&\hspace{1cm}[(w_1+w_2-1)!]^{m_1+m_2+\frac{1-w_1w_2-w_1m_1-w_2m_2}{w_1+w_2-1}}\times x_{12}^{m_1+m_2+\frac{1-w_1w_2-w_1m_1-w_2m_2}{w_1+w_2-1}}\\
&\hspace{1.5cm}\times(\partial_2^{w_1+w_2-1}\gamma)^{-1-\frac{1-w_1w_2-w_1m_1-w_2m_2}{w_1+w_2-1}}\delta_{w_1+w_2-1}(\gamma(z_2)-x_2)+\cdots\,.
\label{eq:numerical prefactor AFTER integration}
\end{aligned}
\end{equation}

\section{Justifying the claim}\label{appendix:proving the claim}
In this appendix, we give a heuristic argument for the claim used in Section \ref{subsec:shorter length}. Let us recall that the claim is that \eqref{eq:OPE expansion in the bosonic string theory} is valid, up to the order shown in the equation, as long as $w_1+w_2-1>0$. In other words, we want to show that, for $w_1+w_2-1>0$,
\begin{equation}
\begin{aligned}
&V^{w_1}_{m_1,j_1,1}(z_1;x_1)V^{w_2}_{m_2,j_2,1}(z_2;x_2)\sim\\
&\hspace{1cm}V^{w_1+w_2-1}_{m_1+m_2+\frac{3}{2},j_1+j_2-\frac{1}{2},1}(z_2;x_2)\delta\left(\frac{x_{12}}{\partial^{w_1+w_2-1}\gamma}-\frac{z_{12}^{w_1+w_2-1}}{(w_1+w_2-1)!}\right)+\cdots\,,
\label{eq:bosonic ads3 OPE}
\end{aligned}
\end{equation}
up to the leading order in the separations $z_{12},x_{12}$. This, together with the OPE \eqref{eq:compact theory OPE} and the usual scaling argument, will allow us to prove that the worldsheet OPE is given by \eqref{eq:OPE expansion in the bosonic string theory} for spectral flows satisfying $w_1+w_2-1>0$, up to numerical factors. Firstly, we note that spectrally flowed operators can be written as a certain line operator acting on unflowed operators, see Section 3 of \cite{Knighton:2023mhq},
\begin{equation}
V^{w_i}_{m_i,j_i,X}(z_i;x_i)=\exp\left(\frac{w_i}{2\pi i}\int_{P_i}dz\ln(z-z_i)({\cal J}^3-x_i{\cal J}^+)\right)V^{0}_{m_i,j_i,X}(z_i;x_i)\,,
\end{equation}
where $P_i$ denotes a keyhole contour around the branch cut of $\ln(z-z_i)$. Thus, we have
\begin{equation}
\begin{aligned}
&V^{w_1}_{m_1,j_1}(z_1;x_1)V^{w_2}_{m_2,j_2}(z_2;x_2)\\
&\hspace{0.7cm}\sim V^{1}_{m_1,j_1}(z_1;x_1)e^{\frac{w_1-1}{2\pi i}\int_{P_1}dz\ln(z-z_1)({\cal J}^3-x_1{\cal J}^+)}e^{\frac{w_2}{2\pi i}\int_{P_2}dz\ln(z-z_2)({\cal J}^3-x_2{\cal J}^+)}V^{0}_{m_2,j_2}(z_2;x_2)\,,
\end{aligned}
\end{equation}
where we have used the notation $V^{w_i}_{m_i,j_i}(z_i;x_i):=V^{w_i}_{m_i,j_i,1}(z_i;x_i)$ to avoid cluttering the expression. Hence, up to the leading order in $z_{12},x_{12}$, we get
\begin{equation}
\begin{aligned}
&e^{\frac{w_1-1}{2\pi i}\int_{P_1}dz\ln(z-z_1)({\cal J}^3-x_1{\cal J}^+)}e^{\frac{w_2}{2\pi i}\int_{P_2}dz\ln(z-z_2)({\cal J}^3-x_2{\cal J}^+)}V^{0}_{m_2,j_2}(z_2;x_2)\\
&= e^{\frac{w_1-1}{2\pi i}\int_{P_1}dz\ln(z-z_2-z_{12})({\cal J}^3-(x_2+x_{12}){\cal J}^+)}e^{\frac{w_2}{2\pi i}\int_{P_2}dz\ln(z-z_2)({\cal J}^3-x_2{\cal J}^+)}V^{0}_{m_2,j_2}(z_2;x_2)\\
&\sim e^{\frac{w_1-1}{2\pi i}\int_{P_2}dz\ln(z-z_2)({\cal J}^3-x_2{\cal J}^+)}e^{\frac{w_2}{2\pi i}\int_{P_2}dz\ln(z-z_2)({\cal J}^3-x_2{\cal J}^+)}V^{0}_{m_2,j_2}(z_2;x_2)\\
&\sim e^{\frac{w_1+w_2-1}{2\pi i}\int_{P_2}dz\ln(z-z_2)({\cal J}^3-x_2{\cal J}^+)}V^{0}_{m_2,j_2}(z_2;x_2)\\
&\sim V^{w_1+w_2-1}_{m_2,j_2}(z_2;x_2)\,.
\end{aligned}
\end{equation}
By our assumption that $w_1+w_2-1>0$, the vertex operator $V^{w_1+w_2-1}_{m_2,j_2}(z_2;x_2)$ takes the form \eqref{eq:bosonic tensionless vertex operator} and therefore, the OPE 
\begin{equation}
V^{1}_{m_1,j_1}(z_1;x_1)V^{w_1+w_2-1}_{m_2,j_2}(z_2;x_2)
\end{equation}
can be found using the approach in Section \ref{subsec:longest length} which gives
\begin{equation}
\begin{aligned}
V^{w_1}_{m_1,j_1}(z_1;x_1)V^{w_2}_{m_2,j_2}(z_2;x_2)\sim& V^{1}_{m_1,j_1}(z_1;x_1)V^{w_1+w_2-1}_{m_2,j_2}(z_2;x_2)\\
\sim&V^{w_1+w_2-1}_{m_1+m_2+\frac{3}{2},j_1+j_2-\frac{1}{2}}(z_2;x_2)\delta\left(\frac{x_{12}}{\partial^{w_1+w_2-1}\gamma}-\frac{z_{12}^{w_1+w_2-1}}{(w_1+w_2-1)!}\right)\,.
\end{aligned}
\end{equation}
Thus, we have justified \eqref{eq:bosonic ads3 OPE}. Combining this with the compact theory $X$ OPE \eqref{eq:compact theory OPE}, we obtain
\begin{equation}
\begin{aligned}
&V^{w_1}_{m_1,j_1,X}(z_1;x_1)V^{w_2}_{m_2,j_2,X}(z_2;x_2)\\
&\sim\sum_aV^{w_1+w_2-1}_{m_1+m_2+\frac{3}{2},j_1+j_2-\frac{1}{2},X_a}(z_2;x_2)\delta\left(\frac{x_{12}}{\partial^{w_1+w_2-1}\gamma}-\frac{z_{12}^{w_1+w_2-1}}{(w_1+w_2-1)!}\right)\\
&\sim\sum_a z_{12}^{-\Delta_1-\Delta_2+\Delta_a+w_1+w_2-1}V^{w_1+w_2-1}_{m_1+m_2+\frac{3}{2},j_1+j_2-\frac{1}{2},X_a}(z_2;x_2)\delta\left(\frac{x_{12}}{\partial^{w_1+w_2-1}\gamma}-\frac{z_{12}^{w_1+w_2-1}}{(w_1+w_2-1)!}\right)\,,
\end{aligned}
\end{equation}
where the $z_{12}$ dependence in the last line was deduced from the standard scaling argument and recall that $\Delta_a$ is the worldsheet conformal weight of $V^{w_1+w_2-1}_{m_1+m_2+\frac{3}{2},j_1+j_2-\frac{1}{2},X_a}(z_2;x_2)$. Integrating over $z_i$ yields
\begin{equation}
\begin{aligned}
&\int dz_1V^{w_1}_{m_1,j_1,X}(z_1;x_1)\int dz_2V^{w_2}_{m_2,j_2,X}(z_2;x_2)\\
&\hspace{1cm}\sim\sum_a x_{12}^{h_a-h_1-h_2}\int dz_2V^{w_1+w_2-1}_{h_a-\frac{3}{2}(w_1+w_2-1),j_1+j_2-\frac{1}{2},X_a}(z_2;x_2)\,,
\end{aligned}
\end{equation}
where we have used that $V^{w_i}_{m_i,j_i,X}(z_i;x_i)$ for $i=1,2$ are on-shell. The OPE above is precisely \eqref{eq:OPE expansion in the bosonic string theory}, however, it is now valid not only for $w_i>0$ but also for $w_1+w_2-1>0$. This purely bosonic analysis can be repeated for hybrid strings and we expect the hybrid string OPEs to be given by the formulas in Section \ref{subsec:longest length} whenever $w_1+w_2-1>0$. In particular, the fusion of spectrally flowed ground states with the screening operator ${\cal O}^-$ can be found using \eqref{eq:tensionless odd x even expansion} and \eqref{eq:tensionless odd x odd expansion}.

\bibliography{bibliography}

\providecommand{\href}[2]{#2}\begingroup\raggedright\begin{thebibliography}{10}

\bibitem{Gaberdiel:2018rqv}
M.~R. Gaberdiel and R.~Gopakumar, ``{Tensionless string spectra on
  AdS$_{3}$},'' \href{http://dx.doi.org/10.1007/JHEP05(2018)085}{{\em JHEP}
  {\bfseries 05} (2018) 085}, \href{http://arxiv.org/abs/1803.04423}{{\ttfamily
  arXiv:1803.04423 [hep-th]}}.

\bibitem{Eberhardt:2018ouy}
L.~Eberhardt, M.~R. Gaberdiel, and R.~Gopakumar, ``{The Worldsheet Dual of the
  Symmetric Product CFT},''
  \href{http://dx.doi.org/10.1007/JHEP04(2019)103}{{\em JHEP} {\bfseries 04}
  (2019) 103}, \href{http://arxiv.org/abs/1812.01007}{{\ttfamily
  arXiv:1812.01007 [hep-th]}}.

\bibitem{Eberhardt:2019ywk}
L.~Eberhardt, M.~R. Gaberdiel, and R.~Gopakumar, ``{Deriving the
  AdS$_{3}$/CFT$_{2}$ correspondence},''
  \href{http://dx.doi.org/10.1007/JHEP02(2020)136}{{\em JHEP} {\bfseries 02}
  (2020) 136}, \href{http://arxiv.org/abs/1911.00378}{{\ttfamily
  arXiv:1911.00378 [hep-th]}}.

\bibitem{Dei:2020zui}
A.~Dei, M.~R. Gaberdiel, R.~Gopakumar, and B.~Knighton, ``{Free field
  world-sheet correlators for ${\rm AdS}_3$},''
  \href{http://dx.doi.org/10.1007/JHEP02(2021)081}{{\em JHEP} {\bfseries 02}
  (2021) 081}, \href{http://arxiv.org/abs/2009.11306}{{\ttfamily
  arXiv:2009.11306 [hep-th]}}.

\bibitem{Giribet:2018ada}
G.~Giribet, C.~Hull, M.~Kleban, M.~Porrati, and E.~Rabinovici, ``{Superstrings
  on AdS$_{3}$ at $k =$ 1},''
  \href{http://dx.doi.org/10.1007/JHEP08(2018)204}{{\em JHEP} {\bfseries 08}
  (2018) 204}, \href{http://arxiv.org/abs/1803.04420}{{\ttfamily
  arXiv:1803.04420 [hep-th]}}.

\bibitem{Gaberdiel:2024dva}
M.~R. Gaberdiel and V.~Sriprachyakul, ``{Tensionless strings on $AdS_3 \times
  S^3 \times S^3 \times S^1$},''
  \href{http://arxiv.org/abs/2411.16848}{{\ttfamily arXiv:2411.16848
  [hep-th]}}.

\bibitem{Dei:2023ivl}
A.~Dei, B.~Knighton, and K.~Naderi, ``{Solving AdS$_{3}$ string theory at
  minimal tension: tree-level correlators},''
  \href{http://dx.doi.org/10.1007/JHEP09(2024)135}{{\em JHEP} {\bfseries 09}
  (2024) 135}, \href{http://arxiv.org/abs/2312.04622}{{\ttfamily
  arXiv:2312.04622 [hep-th]}}.

\bibitem{Gaberdiel:2021kkp}
M.~R. Gaberdiel, B.~Knighton, and J.~Vo\v{s}mera, ``{D-branes in AdS$_{3}$
  \texttimes{} S$^{3}$ \texttimes{} \ensuremath{\mathbb{T}}$^{4}$ at k = 1 and
  their holographic duals},''
  \href{http://dx.doi.org/10.1007/JHEP12(2021)149}{{\em JHEP} {\bfseries 12}
  (2021) 149}, \href{http://arxiv.org/abs/2110.05509}{{\ttfamily
  arXiv:2110.05509 [hep-th]}}.

\bibitem{Knighton:2024noc}
B.~Knighton, V.~Sriprachyakul, and J.~Vo\v{s}mera, ``{Topological defects and
  tensionless holography},'' \href{http://arxiv.org/abs/2406.03467}{{\ttfamily
  arXiv:2406.03467 [hep-th]}}.

\bibitem{Gutperle:2024vyp}
M.~Gutperle, Y.-Y. Li, D.~Rathore, and K.~Roumpedakis, ``{Non-invertible
  symmetries in S$_{N}$ orbifold CFTs and holography},''
  \href{http://dx.doi.org/10.1007/JHEP09(2024)110}{{\em JHEP} {\bfseries 09}
  (2024) 110}, \href{http://arxiv.org/abs/2405.15693}{{\ttfamily
  arXiv:2405.15693 [hep-th]}}.

\bibitem{Harris:2025wak}
S.~Harris, Y.~Hikida, V.~Schomerus, and T.~Tsuda, ``{Holographic Interfaces in
  Symmetric Product Orbifolds},''
  \href{http://arxiv.org/abs/2504.00078}{{\ttfamily arXiv:2504.00078
  [hep-th]}}.

\bibitem{Eberhardt:2021jvj}
L.~Eberhardt, ``{Summing over Geometries in String Theory},''
  \href{http://dx.doi.org/10.1007/JHEP05(2021)233}{{\em JHEP} {\bfseries 05}
  (2021) 233}, \href{http://arxiv.org/abs/2102.12355}{{\ttfamily
  arXiv:2102.12355 [hep-th]}}.

\bibitem{Aharony:2024fid}
O.~Aharony and E.~Y. Urbach, ``{Type II string theory on
  AdS3\texttimes{}S3\texttimes{}T4 and symmetric orbifolds},''
  \href{http://dx.doi.org/10.1103/PhysRevD.110.046028}{{\em Phys. Rev. D}
  {\bfseries 110} no.~4, (2024) 046028},
  \href{http://arxiv.org/abs/2406.14605}{{\ttfamily arXiv:2406.14605
  [hep-th]}}.

\bibitem{Kim:2015gak}
J.~Kim and M.~Porrati, ``{On the central charge of spacetime current algebras
  and correlators in string theory on AdS$_{3}$},''
  \href{http://dx.doi.org/10.1007/JHEP05(2015)076}{{\em JHEP} {\bfseries 05}
  (2015) 076}, \href{http://arxiv.org/abs/1503.07186}{{\ttfamily
  arXiv:1503.07186 [hep-th]}}.

\bibitem{Naderi:2024wqx}
K.~Naderi, ``{Space-time symmetry from the world-sheet},''
  \href{http://dx.doi.org/10.1007/JHEP03(2025)128}{{\em JHEP} {\bfseries 03}
  (2025) 128}, \href{http://arxiv.org/abs/2407.15575}{{\ttfamily
  arXiv:2407.15575 [hep-th]}}.

\bibitem{Eberhardt:2021vsx}
L.~Eberhardt, ``{A perturbative CFT dual for pure NS\textendash{}NS AdS$_{3}$
  strings},'' \href{http://dx.doi.org/10.1088/1751-8121/ac47b2}{{\em J. Phys.
  A} {\bfseries 55} no.~6, (2022) 064001},
  \href{http://arxiv.org/abs/2110.07535}{{\ttfamily arXiv:2110.07535
  [hep-th]}}.

\bibitem{Dei:2022pkr}
A.~Dei and L.~Eberhardt, ``{String correlators on $\text{AdS}_3$: Analytic
  structure and dual CFT},''
  \href{http://dx.doi.org/10.21468/SciPostPhys.13.3.053}{{\em SciPost Phys.}
  {\bfseries 13} no.~3, (2022) 053},
  \href{http://arxiv.org/abs/2203.13264}{{\ttfamily arXiv:2203.13264
  [hep-th]}}.

\bibitem{Knighton:2024qxd}
B.~Knighton and V.~Sriprachyakul, ``{Unravelling AdS$_{3}$/CFT$_{2}$ near the
  boundary},'' \href{http://dx.doi.org/10.1007/JHEP01(2025)042}{{\em JHEP}
  {\bfseries 01} (2025) 042}, \href{http://arxiv.org/abs/2404.07296}{{\ttfamily
  arXiv:2404.07296 [hep-th]}}.

\bibitem{Knighton:2023mhq}
B.~Knighton, S.~Seet, and V.~Sriprachyakul, ``{Spectral flow and localisation
  in AdS$_{3}$ string theory},''
  \href{http://dx.doi.org/10.1007/JHEP05(2024)113}{{\em JHEP} {\bfseries 05}
  (2024) 113}, \href{http://arxiv.org/abs/2312.08429}{{\ttfamily
  arXiv:2312.08429 [hep-th]}}.

\bibitem{Hikida:2023jyc}
Y.~Hikida and V.~Schomerus, ``{Engineering perturbative string duals for
  symmetric product orbifold CFTs},''
  \href{http://dx.doi.org/10.1007/JHEP06(2024)071}{{\em JHEP} {\bfseries 06}
  (2024) 071}, \href{http://arxiv.org/abs/2312.05317}{{\ttfamily
  arXiv:2312.05317 [hep-th]}}.

\bibitem{Sriprachyakul:2024gyl}
V.~Sriprachyakul, ``{Superstrings near the conformal boundary of AdS$_{3}$},''
  \href{http://dx.doi.org/10.1007/JHEP08(2024)203}{{\em JHEP} {\bfseries 08}
  (2024) 203}, \href{http://arxiv.org/abs/2405.03678}{{\ttfamily
  arXiv:2405.03678 [hep-th]}}.

\bibitem{Sriprachyakul:2024xih}
V.~Sriprachyakul, ``{Spacetime dilaton in AdS$_{3}$ \texttimes{} X
  holography},'' \href{http://dx.doi.org/10.1007/JHEP11(2024)083}{{\em JHEP}
  {\bfseries 11} (2024) 083}, \href{http://arxiv.org/abs/2408.13488}{{\ttfamily
  arXiv:2408.13488 [hep-th]}}.

\bibitem{Yu:2024kxr}
Z.-f. Yu and C.~Peng, ``{Correlators of long strings on
  AdS$_{3}$\texttimes{}S$^{3}$\texttimes{}T$^{4}$},''
  \href{http://dx.doi.org/10.1007/JHEP01(2025)017}{{\em JHEP} {\bfseries 01}
  (2025) 017}, \href{http://arxiv.org/abs/2408.16712}{{\ttfamily
  arXiv:2408.16712 [hep-th]}}.

\bibitem{Seiberg:1999xz}
N.~Seiberg and E.~Witten, ``{The D1 / D5 system and singular CFT},''
  \href{http://dx.doi.org/10.1088/1126-6708/1999/04/017}{{\em JHEP} {\bfseries
  04} (1999) 017}, \href{http://arxiv.org/abs/hep-th/9903224}{{\ttfamily
  arXiv:hep-th/9903224}}.

\bibitem{Knighton:2024pqh}
B.~Knighton, ``{Deriving the long-string CFT in AdS$_3$},''
  \href{http://arxiv.org/abs/2410.16904}{{\ttfamily arXiv:2410.16904
  [hep-th]}}.

\bibitem{Balthazar:2021xeh}
B.~Balthazar, A.~Giveon, D.~Kutasov, and E.~J. Martinec, ``{Asymptotically free
  AdS$_{3}$/CFT$_{2}$},'' \href{http://dx.doi.org/10.1007/JHEP01(2022)008}{{\em
  JHEP} {\bfseries 01} (2022) 008},
  \href{http://arxiv.org/abs/2109.00065}{{\ttfamily arXiv:2109.00065
  [hep-th]}}.

\bibitem{Chakraborty:2025nlb}
S.~Chakraborty, A.~Giveon, and D.~Kutasov, ``{Effective AdS$_{3}$/CFT$_{2}$},''
  \href{http://dx.doi.org/10.1007/JHEP03(2025)030}{{\em JHEP} {\bfseries 03}
  (2025) 030}, \href{http://arxiv.org/abs/2501.09119}{{\ttfamily
  arXiv:2501.09119 [hep-th]}}.

\bibitem{Kutasov:1999xu}
D.~Kutasov and N.~Seiberg, ``{More comments on string theory on AdS(3)},''
  \href{http://dx.doi.org/10.1088/1126-6708/1999/04/008}{{\em JHEP} {\bfseries
  04} (1999) 008}, \href{http://arxiv.org/abs/hep-th/9903219}{{\ttfamily
  arXiv:hep-th/9903219}}.

\bibitem{Berkovits:1999im}
N.~Berkovits, C.~Vafa, and E.~Witten, ``{Conformal field theory of AdS
  background with Ramond-Ramond flux},''
  \href{http://dx.doi.org/10.1088/1126-6708/1999/03/018}{{\em JHEP} {\bfseries
  03} (1999) 018}, \href{http://arxiv.org/abs/hep-th/9902098}{{\ttfamily
  arXiv:hep-th/9902098}}.

\bibitem{Maldacena:2000hw}
J.~M. Maldacena and H.~Ooguri, ``{Strings in AdS(3) and SL(2,R) WZW model 1.:
  The Spectrum},'' \href{http://dx.doi.org/10.1063/1.1377273}{{\em J. Math.
  Phys.} {\bfseries 42} (2001) 2929--2960},
  \href{http://arxiv.org/abs/hep-th/0001053}{{\ttfamily arXiv:hep-th/0001053}}.

\bibitem{Maldacena:2000kv}
J.~M. Maldacena, H.~Ooguri, and J.~Son, ``{Strings in AdS(3) and the SL(2,R)
  WZW model. Part 2. Euclidean black hole},''
  \href{http://dx.doi.org/10.1063/1.1377039}{{\em J. Math. Phys.} {\bfseries
  42} (2001) 2961--2977}, \href{http://arxiv.org/abs/hep-th/0005183}{{\ttfamily
  arXiv:hep-th/0005183}}.

\bibitem{Maldacena:2001km}
J.~M. Maldacena and H.~Ooguri, ``{Strings in AdS(3) and the SL(2,R) WZW model.
  Part 3. Correlation functions},''
  \href{http://dx.doi.org/10.1103/PhysRevD.65.106006}{{\em Phys. Rev. D}
  {\bfseries 65} (2002) 106006},
  \href{http://arxiv.org/abs/hep-th/0111180}{{\ttfamily arXiv:hep-th/0111180}}.

\bibitem{Henningson:1991jc}
M.~Henningson, S.~Hwang, P.~Roberts, and B.~Sundborg, ``{Modular invariance of
  SU(1,1) strings},''
  \href{http://dx.doi.org/10.1016/0370-2693(91)90944-L}{{\em Phys. Lett. B}
  {\bfseries 267} (1991) 350--355}.

\bibitem{Hwang:1991ana}
S.~Hwang, ``{Cosets as gauge slices in SU(1,1) strings},''
  \href{http://dx.doi.org/10.1016/0370-2693(92)91666-W}{{\em Phys. Lett. B}
  {\bfseries 276} (1992) 451--454},
  \href{http://arxiv.org/abs/hep-th/9110039}{{\ttfamily arXiv:hep-th/9110039}}.

\bibitem{Evans:1998qu}
J.~M. Evans, M.~R. Gaberdiel, and M.~J. Perry, ``{The no ghost theorem for
  AdS(3) and the stringy exclusion principle},''
  \href{http://dx.doi.org/10.1016/S0550-3213(98)00561-6}{{\em Nucl. Phys. B}
  {\bfseries 535} (1998) 152--170},
  \href{http://arxiv.org/abs/hep-th/9806024}{{\ttfamily arXiv:hep-th/9806024}}.

\bibitem{Giveon:1998ns}
A.~Giveon, D.~Kutasov, and N.~Seiberg, ``{Comments on string theory on
  AdS(3)},'' \href{http://dx.doi.org/10.4310/ATMP.1998.v2.n4.a3}{{\em Adv.
  Theor. Math. Phys.} {\bfseries 2} (1998) 733--782},
  \href{http://arxiv.org/abs/hep-th/9806194}{{\ttfamily arXiv:hep-th/9806194}}.

\bibitem{deBoer:1998gyt}
J.~de~Boer, H.~Ooguri, H.~Robins, and J.~Tannenhauser, ``{String theory on
  AdS(3)},'' \href{http://dx.doi.org/10.1088/1126-6708/1998/12/026}{{\em JHEP}
  {\bfseries 12} (1998) 026},
  \href{http://arxiv.org/abs/hep-th/9812046}{{\ttfamily arXiv:hep-th/9812046}}.

\bibitem{Wakimoto:1986gf}
M.~Wakimoto, ``{Fock representations of the affine lie algebra A1(1)},''
  \href{http://dx.doi.org/10.1007/BF01211068}{{\em Commun. Math. Phys.}
  {\bfseries 104} (1986) 605--609}.

\bibitem{Naderi:2022bus}
K.~Naderi, ``{DDF operators in the hybrid formalism},''
  \href{http://dx.doi.org/10.1007/JHEP12(2022)043}{{\em JHEP} {\bfseries 12}
  (2022) 043}, \href{http://arxiv.org/abs/2208.01617}{{\ttfamily
  arXiv:2208.01617 [hep-th]}}.

\bibitem{Fiset:2022erp}
M.-A. Fiset, M.~R. Gaberdiel, K.~Naderi, and V.~Sriprachyakul, ``{Perturbing
  the symmetric orbifold from the worldsheet},''
  \href{http://dx.doi.org/10.1007/JHEP07(2023)093}{{\em JHEP} {\bfseries 07}
  (2023) 093}, \href{http://arxiv.org/abs/2212.12342}{{\ttfamily
  arXiv:2212.12342 [hep-th]}}.

\bibitem{Jevicki:1998bm}
A.~Jevicki, M.~Mihailescu, and S.~Ramgoolam, ``{Gravity from CFT on S**N(X):
  Symmetries and interactions},''
  \href{http://dx.doi.org/10.1016/S0550-3213(00)00147-4}{{\em Nucl. Phys. B}
  {\bfseries 577} (2000) 47--72},
  \href{http://arxiv.org/abs/hep-th/9907144}{{\ttfamily arXiv:hep-th/9907144}}.

\bibitem{Pakman:2009zz}
A.~Pakman, L.~Rastelli, and S.~S. Razamat, ``{Diagrams for Symmetric Product
  Orbifolds},'' \href{http://dx.doi.org/10.1088/1126-6708/2009/10/034}{{\em
  JHEP} {\bfseries 10} (2009) 034},
  \href{http://arxiv.org/abs/0905.3448}{{\ttfamily arXiv:0905.3448 [hep-th]}}.

\bibitem{Burrington:2018upk}
B.~A. Burrington, I.~T. Jardine, and A.~W. Peet, ``{The OPE of bare twist
  operators in bosonic $S_N$ orbifold CFTs at large $N$},''
  \href{http://dx.doi.org/10.1007/JHEP08(2018)202}{{\em JHEP} {\bfseries 08}
  (2018) 202}, \href{http://arxiv.org/abs/1804.01562}{{\ttfamily
  arXiv:1804.01562 [hep-th]}}.

\bibitem{DeBeer:2019oxm}
T.~De~Beer, B.~A. Burrington, I.~T. Jardine, and A.~W. Peet, ``{The large $N$
  limit of OPEs in symmetric orbifold CFTs with $\mathcal{N}=(4,4)$
  supersymmetry},'' \href{http://dx.doi.org/10.1007/s13130-019-11019-2}{{\em
  JHEP} {\bfseries 08} (2019) 015},
  \href{http://arxiv.org/abs/1904.07816}{{\ttfamily arXiv:1904.07816
  [hep-th]}}.

\bibitem{Ashok:2023kkd}
S.~K. Ashok and J.~Troost, ``{The operator rings of topological symmetric
  orbifolds and their large N limit},''
  \href{http://dx.doi.org/10.1007/JHEP04(2024)039}{{\em JHEP} {\bfseries 04}
  (2024) 039}, \href{http://arxiv.org/abs/2309.17052}{{\ttfamily
  arXiv:2309.17052 [hep-th]}}.

\bibitem{Lunin:2001pw}
O.~Lunin and S.~D. Mathur, ``{Three point functions for M(N) / S(N) orbifolds
  with N=4 supersymmetry},''
  \href{http://dx.doi.org/10.1007/s002200200638}{{\em Commun. Math. Phys.}
  {\bfseries 227} (2002) 385--419},
  \href{http://arxiv.org/abs/hep-th/0103169}{{\ttfamily arXiv:hep-th/0103169}}.

\bibitem{Lunin:2000yv}
O.~Lunin and S.~D. Mathur, ``{Correlation functions for M**N / S(N)
  orbifolds},'' \href{http://dx.doi.org/10.1007/s002200100431}{{\em Commun.
  Math. Phys.} {\bfseries 219} (2001) 399--442},
  \href{http://arxiv.org/abs/hep-th/0006196}{{\ttfamily arXiv:hep-th/0006196}}.

\bibitem{Polchinski:1998rq}
J.~Polchinski, \href{http://dx.doi.org/10.1017/CBO9780511816079}{{\em {String
  theory. Vol. 1: An introduction to the bosonic string}}}.
\newblock Cambridge Monographs on Mathematical Physics. Cambridge University
  Press, 12, 2007.

\bibitem{Gaberdiel:2022als}
M.~R. Gaberdiel, K.~Naderi, and V.~Sriprachyakul, ``{The free field realisation
  of the BVW string},'' \href{http://dx.doi.org/10.1007/JHEP08(2022)274}{{\em
  JHEP} {\bfseries 08} (2022) 274},
  \href{http://arxiv.org/abs/2202.11392}{{\ttfamily arXiv:2202.11392
  [hep-th]}}.

\end{thebibliography}\endgroup
\bibliographystyle{utphys.sty}

\end{document}